\documentclass[twoside]{reporteq}

\usepackage{amsmath}
\usepackage{amsfonts}
\usepackage{amssymb}
\usepackage{latexsym}
\usepackage{times}     
\usepackage{pstricks}
\usepackage{pst-node}
\usepackage{svcon2e}
\usepackage{makeidx}

\def\date{August 12, 2002}
\psset{framearc=0.0,framesep=0.0truecm,nodesep=3pt,unit=0.75truecm,%
arrowsize=4pt 2,arrowinset=0.4}
\def\pstlw{0.8pt}                              
\newcommand{\ed}{\end{document}}
\makeindex
\renewcommand{\theequation}{\arabic{section}.\arabic{equation}}
\newcounter{mycnt}[section]
\def\themycnt{\addtocounter{mycnt}{1}\thechapter.\thesection.\arabic{mycnt}}
\def\mybenv#1{\vskip 3pt\noindent{\bf #1 \themycnt}\vskip 3pt\indent}
\def\myeenv{\hfill\rule{1ex}{1ex}\vskip 3pt}
\newcommand{\JJ}{\,\raisebox{0.33ex}{\rule{1ex}{1pt}\rule{1pt}{1ex}}}
\newcommand{\LL}{\,\raisebox{0.33ex}{\rule{1pt}{1ex}\rule{1ex}{1pt}}}
\def\length#1{\text{length}(#1)}
\def\nn{\nonumber \\}
\def\Id{\text{I\!d}}
\def\openk{\Bbbk}
\def\openC{\mathbb{C}}

\def\openZ{\mathbb{Z}}

\def\!{\kern -0.15ex}
\def\nn{\nonumber\\}
\begin{document}
\chapter{Grade free product formul\ae\ from Gra{\ss}mann Hopf gebras}
\chapterauthors{Bertfried Fauser}
{ {\renewcommand{\thefootnote}{\fnsymbol{footnote}}
\footnotetext{\kern-15.3pt AMS Subject Classification: 16W30; 15A66.}
}}
\begin{abstract}
In the traditional approaches to Clifford algebras, the Clifford product
is evaluated by recursive application of the product of a one-vector (span
of the generators) on homogeneous i.e. sums of decomposable (Gra{\ss}mann),
multi-vectors and later extended by bilinearity. The Hestenesian 'dot'
product, extending the one-vector scalar product, is even worse having
exceptions for scalars and the need for applying grade operators at
various times. Moreover, the multivector grade is not a generic Clifford
algebra concept. The situation becomes even worse in geometric
applications if a meet, join or contractions have to be calculated.  
 
Starting from a naturally graded Gra{\ss}mann Hopf gebra, we derive general
formul\ae\ for the products: meet and join, comeet and cojoin, left/right
contraction, left/right cocontraction, Clifford and co-Clifford products. 
All these product formul\ae\ are valid for any grade and any inhomogeneous
multivector factors in Clifford algebras of any bilinear form, including
non-symmetric and degenerated forms. We derive the three well known
Chevalley formul\ae\ as a specialization of our approach and will
display co-Chevalley formul\ae. The Rota--Stein cliffordization is
shown to be the generalization of Chevalley deformation. Our product
formul\ae\ are based on invariant theory and are not tied to
representations/matrices and are highly computationally effective. The
method is applicable to symplectic Clifford algebras too.      

\noindent 
{\bf Keywords: } Gra{\ss}mann Hopf gebra, contraction, cocontraction,
Chevalley deformation, Rota--Stein cliffordization, Clifford product, 
Clifford coproduct, meet, join, comeet, cojoin, contractions, 
cocontractions, linear duality, categorial duality, Gra{\ss}mann-Cayley
algebra 
\end{abstract}
\section{Introduction}

\subsection{Preliminary note}

Beside some rumour during the conference, we continue to use 
algebra, cogebra and Hopf gebra as technical terms. In our eyes these 
names fit into mathematical nomenclature having also a linguistic
background. The most striking argument is, however, that it is 
misleading to call a cogebra a co{\em al}gebra making use of and
pointing to the term {\em algebra}. By duality one sees that cogebras 
contain in principle the same amount of information as algebras. One 
could (should?) come up with a linear cogebra theory not making use of 
any algebraic structure or knowledge. It seems necessary to us to put
the finger into the wound of the missed opportunity \cite{dyson:1974a}
to develop algebra and coalgebra on the same footing and beg for pardon to 
those who feel linguistically offended by our naming.
\index{cogebra!naming}
\index{cogebra!versus co{\em al}gebra}

\subsection{Synopsis}

The present paper will gather {\em grade free} product formul\ae\ for 
almost all algebra and cogebra products related to Gra{\ss}mann-,
Gra{\ss}mann-Cayley, and Clifford algebras. This does not mean that we
abandon the multivector structures of these algebras but that we come 
up with formul\ae\ which are valid for general multivector polynomials, 
i.e. for general elements $x$ from the algebra $A$ or cogebra $C$. In 
present literature important product formul\ae\ are given only on 
generators or homogeneous elements of certain grades, and have to be 
expanded by iteration and linearity to the general case. Among these the 
most important Clifford product has to be calculated this way!

\index{product!with grade restrictions}
In \cite{hestenes:sobczyk:1992a} we find formul\ae\ 
(1.21a-c), (1.22a,c), (1.23a,b), (1.25b,c) etc. where even restrictions 
like that the grade of one homogeneous algebra element has to be less or
equal to the grade of another such element, e.g. (1.23a) and (1.25b,c),
have to be assumed. The situation even goes worse if dot and inner 
products are considered. It is not our aim to criticise but to overcome 
this deficiencies. During this course we will gain lots of insights
into the (almost) perfectly dual structure of algebras and cogebras.

To reach our goal we will see that we {\em have to} employ algebra and cogebra 
structures. Furthermore we will take as our point of departure the well 
behaved Gra{\ss}mann Hopf algebra. Firstly we will show that the 
Gra{\ss}mann-Cayley algebra is related to the Gra{\ss}mann Hopf algebra 
by dualizing the coproduct. Then by deformation we will reach contractions
and Clifford algebras. It will turn out that the Chevalley deformation 
having a grade restriction is a particular case of the general Rota-Stein 
{\em cliffordization} obeying no grade restriction.

Using categorial duality, we can write down immediately dualized versions 
of all algebraic well know structures coming up with a self dual 
Gra{\ss}mann-Cayley double algebra, cocontractions, Chevalley codeformation,
and with Clifford cogebras etc. 

Categorial duality employs a most powerful and beautiful symmetry. To 
reach our results {\em cogebra structures} are inevitable. We strongly
belief that only a fully dual treatment of (projective) geometry, 
linear co/al-gebra, invariant and deformation theory will prove powerful
enough to overcome recent problems in mathematics and physics.

\subsection{The grading}

Since in various discussions, which took place during the ICCA 6,
it became clear to me that the concept of grading and filtration
seems not to be common ground, we will first settle down this issue here.

\index{module!graded}
\index{grading}
A {\em graded $\openk$-module} $A$ (graded $\openk$-vector space, 
or simply {\em linear space}) is a (finite) family of $\openk$-modules 
$\{A_n\}$ where $n$ runs through the non-negative integers. $n$ is 
called {\em degree} or step, grade etc. The degree of an element $a$ is 
denoted in various ways, 
$\partial a = \vert a \vert = \text{length}(a) = \text{deg}(a) = \ldots$.
Let $A,B$ be graded $\openk$-modules. A {\em graded morphism} $f$ 
is a family of morphisms $\{f_i\}$ such that the 
$f_i : A_i \rightarrow B_i$ are morphisms of $\openk$-modules.
An element $a \in A$ is called {\em homogeneous} of degree $r$ iff
one has $a \in A_i$ and $a \not\in A_j, i\not=j$. Any module can be 
trivially graded by declaring its degree to be zero.
\index{grade}
\index{length|see{grade}}
\index{degree|see{grade}}
 
A grading can be introduced also by the action of an abelian group $G$
such that the modules $A_i$ are invariant subspaces of the group
labelled by a representation index (character): $G \bullet A_i \subset A_i$, 
$\chi_j(A_i)=i\delta_{i,j}$.

\index{filtration}
\index{product!filtered}
A {\em filtration} is defined in analogous way demanding the weaker 
obstruction to modules, morphisms etc. that they consists of or map 
into the spaces of same or lower degree
\begin{align}
f_i : A_i \rightarrow \oplus_{j\le i} B_j.
\end{align} 
We will later note that products emerging from cliffordization will be
in general not graded morphisms but only obey a filtration.

\index{homogeneous element}
\mybenv{Example:}
Consider a polynom $p(x)=\alpha_0 + \alpha_1 x + \alpha_2 x^2 + \ldots 
\,\in \openk[[x]]$ in one variable $x$ over the ring (field) $\openk$. 
The 1-dimensional spaces spanned by $x^i$ are $\openk$-modules $A_i$. 
Monoms $\alpha\,x^q$ are homogenous elements of degree $q$. Polynoms in 
several commuting complex variables $\openC[[z,w]]$ can be graded by their 
{\em total degree}: 
$p(z,w) = \alpha_0 +\alpha_{1,0}z +\alpha_{0,1}w
+\alpha_{2,0}z^2 +\alpha_{1,1}zw +\alpha_{0,2}w^2 +\ldots$. The 
$\openC$-spaces of degree $q$ have dimensions $q+1$. Observe that 
one could introduce a finer grading by specifying a {\em multidegree}
composed from the degree in $z$ and $w$, e.g. degree $\partial (z^4w^3)
= \partial(z^4) + \partial(w^3) = 4+3=7$, multidegree $\partial (z^4w^3)
= (4,3)$.
\myeenv

A binary {\em product} $m$ (a binary multiplication) is a morphism from 
the space $B \simeq A\otimes A$ into the space $A$. If $A$ is a graded 
space we can define the grading of $A\otimes A$ to be the sum of the 
grades of the homogeneous factors, i.e. 
$\partial (A_i \otimes A_j) = \partial A_i + \partial A_j$, turning
$B$ into a graded module. A product $m : B \rightarrow A$ is graded 
if it is a graded  morphism. In other terms
\index{product!graded}
\index{product!binary}
\index{product!as morphism}
\begin{align}
m : A_i \otimes A_j &\subset A_{i+j}
\end{align}
Note that this definition of a product implies bilinearity but not 
associativity. We denote the product by $m(a\otimes b) = m(a,b)$, or
in infix notation or even by juxtaposition $a\, m\, b = ab$. Let 
$\alpha, \beta$ be ring elements we have right and left 
{\em distributive laws} (linearity)
\index{product!linearity of}
\index{product!distributive}
\begin{align}
m(\alpha\,a+\beta\,b,c) &= \alpha\,m(a,c) + \beta\,m(b,c) \nn
m(a,\alpha\,b+\beta\,c) &= \alpha\,m(a,b) + \beta\,m(a,c)
\end{align}
If a product acts on two adjacent slots of a higher tensor space 
(two out of a larger number of arguments) it is easily proven to be 
multilinear. Since we deal with associative products mainly, we assume 
$m$ to be associative from now on. In this case we can define
$m(a\otimes \ldots \otimes b) = m(a,\ldots,b) =m(a,m(\ldots))$
where the order of binary multiplications is irrelevant.
\index{associativity}

\mybenv{Example:}
Canonical examples of graded binary products are tensor products, 
Gra{\ss}mann and symmetric products. Let $A=a\ldots b$, $B=c\ldots d$,
be two words in a tensor algebra $T(V)$ generated by the letters 
$a,b,\ldots \in V$ linear over some ring $\openk$. A grading is define 
by the length of the words $\text{length}(A)=r$, $\text{length}(B)=s$ 
the tensor product is concatenation. We find
\begin{align}
m(A\otimes B) &=AB  &&& \length{m(A \otimes B)}
=\length{A}+\length{B}
\end{align}
This allows to decompose the tensor algebra, viewed as module, into a 
sum of disjoint submodules containing homogeneous elements 
$T(V) =\openk \oplus V \oplus V^{\otimes^2}\oplus \ldots$ 
The elements of $V^{\otimes^r}$ need not to be decomposable (products
of generators) but may be sums of products of generators.

\index{generators}
Let $e_1,e_2,\ldots$ be {\em generators} of a Gra{\ss}mann algebra
$V^{\wedge}= \openk\oplus V \oplus V^{\wedge^2} + \ldots$. A grading
can be defined by the number of generators in a monomial. If we define 
$0$ to have any grade, we find that the Gra{\ss}mann wedge product 
is a graded product
\begin{align}
&V^{\wedge^r} \wedge V^{\wedge^s} \subset V^{\wedge^{r+s}} \nn
&\text{length}(V^{\wedge^r} \wedge V^{\wedge^s})
=\length{V^{\wedge^r}}+\length{V^{\wedge^s}} = r+s.
\end{align}

A symmetric product in $\openk[[a,b,\ldots]]$ defined as the usual 
point wise product of polynomials is graded. 
\begin{align}
m(a^3b^2\otimes (c^2d+d^3)) &=a^3b^2c^2d+a^3b^2d^3  \\
\length{m(a^3b^2\otimes (c^2d+d^3)}
&=\length{a^3b^2}+\length{c^2d+d^3}= 5+3.\nonumber
\end{align} 
\myeenv

It is possible to derive algebras from the tensor algebra by factoring 
out bilateral ideals. These ideals are generated by elements fulfilling
some relations. In the case of the Gra{\ss}mann and symmetric algebras 
they read for $x,y \in V$
\index{ideal!bilateral}
\begin{align}
{\cal I}_{Gr} &= \text{gen}\{ x\otimes y + y\otimes x\} \nn
{\cal I}_{sym} &= \text{gen}\{ x\otimes y - y \otimes x\}
\end{align}
Since these ideals are graded, the factored algebras remain to be graded 
by the $\openZ$-grading inherited from the tensor algebra. Without going 
into detail of this construction, we find immediately that the ideal
of a Clifford algebra generated as follows
\index{grading!multivector $\openZ$}
\index{grading!parity $\openZ_2$}
\begin{align}
{\cal I}_{Cl} &= \text{gen}\{ x\otimes y +y\otimes x - 2g(x,y)\Id \},
\end{align}
where $g(x,y)$ is the symmetric polar bilinear form of a quadratic form
$Q$ on $V$, is {\em no longer $\openZ$-graded} since tensors of different
degree are identified.
\index{bilinear form!polar}

As a good example to this claim and a counter example to the
widely accepted assumption that a Clifford algebra comes up with 
generic `multivectors' i.e. a $\openZ$-grading may serve the quaternions.
\mybenv{Example:}
Let ${\bf 1,i,j,k}$ be the standard basis of the quaternions, obeying
the relations
\index{grading!by length}
\index{grading!of quaternions}
\index{grading!counter example}
\begin{align}
& {\bf k} = {\bf i}{\bf j} \nn
&{\bf i}{\bf j} = - {\bf j}{\bf i}\qquad
 {\bf j}{\bf k} = - {\bf k}{\bf j}\qquad
 {\bf k}{\bf i} = - {\bf i}{\bf k} \nn
&{\bf i}{\bf i}={\bf j}{\bf j}={\bf k}{\bf k}=
 {\bf i}{\bf j}{\bf k}=-{\bf 1}
\end{align} 
We obtain a grading using the following length function. Assume that
${\bf i}$, ${\bf j}$ are generators and define
\begin{align}
& \length{{\bf 1}}=0&&\length{{\bf i}}=1&&\length{{\bf j}}=1
&&\length{{\bf k}}=2.&
\end{align}
However, the roles of ${\bf i,j,k}$ are fully symmetric and we could 
have chosen that ${\bf j}$, ${\bf k}$ are generators so that 
${\bf i} = {\bf j}{\bf k}$ which would have lead us to a second 
different $\openZ$-grading
\begin{align}
& \length{{\bf 1}}=0&&\length{{\bf j}}=1&&\length{{\bf k}}=1
&&\length{{\bf i}}=2.&
\end{align}
Hence there is no unique such grading present in the quaternions. The
argument above using the tensor algebra and factorization shows that 
such a grading cannot uniquely be established in any Clifford algebra. 
Only the $\openZ_2$-grading or {\em parity grading} defined by the length 
function modulo 2 is generic. 
\myeenv 

Adding a multivector structure to a Clifford algebra depends on
additional choices, e.g. by the choice of particular elements 
being generators. In fact one has to choose in which way a Gra{\ss}mann
algebra having multivectors is embedded in a Clifford algebra.
We are consequently using such an identification in the present work
and all gradings we refer to are derived from the grading of the
tensor and Gra{\ss}mann algebras.

\subsection{Algebra and cogebra}

We will informally introduce the notion of a cogebra by dualizing
the algebra structure. In category theory one uses {\em commutative
diagrams} (CD) for this purpose, however, we also make frequent use of 
{\em tangles}, see discussion and references in \cite{fauser:2002c}. 
The difference between both pictures is that they are dual in the 
sense that arrows and objects change their graphical representation. 
A product $m$ may be seen as a morphisms (arrow) acting 
on objects  (source and target points) in a CD. In the tangle analog 
we represent morphisms by points and objects by lines (arrows, implicitly 
red downwards unless otherwise specified). Categorial duality is the 
operation which reverses all arrows or mirrors all tangles at a horizontal 
line. The therefrom generated dualized morphisms are named using the prefix
`{\em co}', e.g. a product changes into a {\em coproduct}. In graphical 
notation we get
\index{duality!categorial}
\index{tangle!categorial duality}
\begin{align}
\begin{array}{c}
\Rnode{AA}{A\otimes A}\\[8ex]
\Rnode{A}{A}
\ncline{->}{AA}{A}
\Aput{m}
\end{array}
\hskip 0.25truecm\cong\hskip 0.25truecm
\pspicture[0.5](0,0)(1,1.5)
\psset{linewidth=\pstlw,xunit=0.5,yunit=0.5,runit=0.5}
\psset{arrowsize=4pt 2,arrowinset=0.4} 
\psline{-}(0,3)(0,2)
\psline{-}(2,3)(2,2)
\psarc(1,2){1.0}{180}{360}
\psline{-}(1,1)(1,0)
\pscircle[linewidth=0.4pt,fillstyle=solid,fillcolor=white](1,1){0.2}   
\rput(1,1.75){$m$}
\endpspicture
\qquad\Leftarrow\,\text{duality}\,\Rightarrow\qquad
\begin{array}{c}
\Rnode{CC}{C}\\[8ex]
\Rnode{C}{C\otimes C}
\ncline{->}{CC}{C}
\Aput{\Delta}
\end{array}
\hskip 0.25truecm\cong\hskip 0.25truecm
\pspicture[0.5](0,0)(1,1.5)
\psset{linewidth=\pstlw,xunit=0.5,yunit=0.5,runit=0.5}
\psset{arrowsize=2pt 2,arrowinset=0.2} 
\psline{-}(0,0)(0,1)
\psline{-}(2,0)(2,1)
\psarc(1,1){1.0}{0}{180}
\psline{-}(1,2)(1,3)
\pscircle[linewidth=0.4pt,fillstyle=solid,fillcolor=white](1,2){0.2}   
\rput(1,1){$\Delta$}
\endpspicture
\end{align}
\index{graphical calculus|see{tangle}}
\index{duality!in graphical calculus}
Tangles can be red like processes in physics, e.g. think of Feynman diagrams,
or flow diagrams in computer science. Elements or spaces enter at the top 
flow down and suffer at the vertices, representing morphisms, some action. 
A binary product combines two inputs into one output, while a binary 
coproduct has one input and two outputs. Such a representation is called
{\em graphical calculus}. Some details and references may be found in 
\cite{fauser:2002c}. If one calculates with tangles an equality is sometimes
called a {\em move}. The coproduct is a $1 \rightarrow 2$ map algebraically 
given as
\index{move}\index{graphical calculus!move} 
\index{coproduct}
\index{product!co|see{coproduct}}
\begin{align}
\Delta ~&:~ C \rightarrow C\otimes C.
\end{align}
The coproduct is in general an indecomposable tensor. It is very
convenient to introduce the Sweedler notation \cite{sweedler:1969a}
\index{Sweedler notation}
\begin{align}
\Delta(x) &= \sum_{r} a_r \otimes b_r 
= \sum_{(x)} x_{(1)} \otimes x_{(2)}
= x_{(1)} \otimes x_{(2)}\nn
\Delta(x^i) &= \sum_{i} \Delta^{jk}_{i}\, x^i
= \sum_{(r)} a^j_{(r)} \otimes b^k_{(r)}
\qquad\text{w.r.t. an arb. basis} 
\end{align}
The $\Delta_i^{jk}$ are called {\em section coefficients}, these constitute
a sort of {\em comultiplication table}. Associativity dualizes to 
coassociativity and its axiom reads as CD or tangle: 
\index{coproduct!multiplication table}
\index{section coefficients}
\index{coassociativity}
\index{tangle!of coassociativity}
\begin{align}
\label{eqn:coass}
\hskip -2.0truecm
\begin{array}{c@{\hskip 1.25truecm}c}
\Rnode{a}{C} & \Rnode{b}{C \otimes C} \\[6ex]
\Rnode{c}{C\otimes C} & \Rnode{d}{C\otimes C\otimes C}
\end{array}
\ncline{->}{a}{b}
\Aput{\Delta}
\ncline{->}{b}{d}
\Aput{\Id \otimes \Delta}
\ncline{->}{a}{c}
\Bput{\Delta}
\ncline{->}{c}{d}
\Aput{\Delta \otimes \Id}
\qquad\qquad
\pspicture[0.3](0,0)(1.0,3.0)
\psset{linewidth=\pstlw,xunit=0.5,yunit=0.5,runit=0.5}
\psset{arrowsize=1pt 2,arrowinset=0.1}
\psline(0,0)(0,1)
\psline(2,0)(2,1)
\psline(3,0)(3,3)
\psarc(1,1){1}{360}{180}
\psline(1,2)(1,3)
\psarc(2,3){1}{0}{180}
\psline(2,4)(2,5)
\pscircle[linewidth=0.4pt,fillstyle=solid,fillcolor=white](1,2){0.2}
\pscircle[linewidth=0.4pt,fillstyle=solid,fillcolor=white](2,4){0.2}
\rput(1,1.25){$\Delta$}
\rput(2,3.25){$\Delta$}
\endpspicture
\qquad=\quad
\pspicture[0.3](0,0)(0.5,3.0)
\psset{linewidth=\pstlw,xunit=0.5,yunit=0.5,runit=0.5}
\psset{arrowsize=1pt 2,arrowinset=0.1}
\psline(0,0)(0,3)
\psline(1,0)(1,1)
\psline(3,0)(3,1)
\psarc(2,1){1}{0}{180}
\psline(2,2)(2,3)
\psarc(1,3){1}{0}{180}
\psline(1,4)(1,5)
\pscircle[linewidth=0.4pt,fillstyle=solid,fillcolor=white](1,4){0.2}
\pscircle[linewidth=0.4pt,fillstyle=solid,fillcolor=white](2,2){0.2}
\rput(1,3.25){$\Delta$}
\rput(2,1.25){$\Delta$}
\endpspicture
\end{align}
\index{counit}
\index{tangle!of counit}
A coproduct may have a {\em counit} which is defined once more by 
dualizing the axioms of the unit. We find $(\epsilon \otimes \Id) 
\Delta = \Id = (\Id \otimes \epsilon)\Delta$ or graphically
\begin{align}
\label{eqn:counit}
\hskip -0.25truecm
\begin{array}{c@{\hskip 1.25truecm}c@{\hskip 1.25truecm}c}
\Rnode{a}{\openk\otimes C} & \Rnode{b}{C\otimes C} & 
\Rnode{c}{C\otimes\openk} \\[6ex]
 & \Rnode{d}{C} &
\ncline{->}{b}{a}
\Bput{\epsilon\otimes \Id}
\ncline{->}{b}{c}
\Aput{\Id\otimes \epsilon}
\ncline{->}{d}{b}
\Aput{\Delta}
\ncline{->}{a}{d}
\Bput{\approx}
\ncline{->}{c}{d}
\Aput{\approx}
\end{array}
\quad
\pspicture[0.4](0,-0.75)(0.5,2.25)
\psset{linewidth=\pstlw,xunit=0.5,yunit=0.5,runit=0.5}
\psset{arrowsize=2pt 2,arrowinset=0.2}
\psline(2,0)(2,1)
\psarc(1,1){1.0}{0}{180}
\psline(1,2)(1,3)
\pscircle[linewidth=0.4pt,fillstyle=solid,fillcolor=white](0,1){0.2}
\pscircle[linewidth=0.4pt,fillstyle=solid,fillcolor=white](1,2){0.2}
\rput(1,1.25){$\Delta$}
\rput(0,0.25){$\epsilon$}
\endpspicture
\quad\quad=\!\!
\pspicture[0.4](0,-0.75)(0.5,2.25)
\psset{linewidth=\pstlw,xunit=0.5,yunit=0.5,runit=0.5}
\psset{arrowsize=2pt 2,arrowinset=0.2}
\psline(1,3)(1,0)
\endpspicture
\,\,=\quad
\pspicture[0.4](0,-0.75)(0.5,2.25)
\psset{linewidth=\pstlw,xunit=0.5,yunit=0.5,runit=0.5}
\psset{arrowsize=2pt 2,arrowinset=0.2}
\psline(0,0)(0,1)
\psarc(1,1){1.0}{0}{180}
\psline(1,2)(1,3)
\pscircle[linewidth=0.4pt,fillstyle=solid,fillcolor=white](2,1){0.2}
\pscircle[linewidth=0.4pt,fillstyle=solid,fillcolor=white](1,2){0.2}
\rput(1,1.25){$\Delta$}
\rput(2,0.25){$\epsilon$}
\endpspicture
\end{align}
The pair ${\cal A} = (A,m)$ is called an (associative possibly 
unital) algebra and the dualized structure ${\cal C} = (C,\Delta)$ 
is called a (coassociative possibly counital) {\em cogebra}. 
\index{algebra!definition of}
\index{cogebra!definition of}
\index{coalgebra|see{cogebra}}

\subsection{Linear duality}

\index{duality!linear}
Since we have already used categorial duality we have a need to 
introduce the technical term {\em linear duality} for the conventional 
dual. Any possibly graded finite dimensional $\openk$-module $A$ comes 
naturally, i.e. functorially, with a {\em linear dual} 
$A^* \simeq \text{lin-hom}(A,\openk)$. Elements $\omega$ of $A^*$ are 
called linear forms. We will freely use the notations
\index{tangle!oriented}
\begin{align}
\omega(x) = \langle \omega \mid x \rangle = \text{eval}(\omega \otimes x)
\end{align}
Arrows are used to indicate the type of the space. Downwards oriented lines 
represent the space $A$ while upward oriented lines depict the dual
space $A^*$. The action of a linear dual on a space is called 
{\em evaluation map}, denoted as $\text{eval}$, due to symmetry we can 
define the action the opposite way around also, thereby identifying $A$ 
with the double dual $A^{**}$. In terms of tangles we write:
\index{evaluation map}
\index{coevaluation map}
\index{tangle!of evaluation}
\index{tangle!of coevaluation}
\begin{align}
\label{eqn:eval}
\pspicture[0.5](0,0)(1,1.5)
\psset{linewidth=\pstlw,xunit=0.5,yunit=0.5,runit=0.5}
\psset{arrowsize=4pt 2,arrowinset=0.4}
\psline(0,3)(0,2)
\psline(2,3)(2,2)
\psline{<-}(0,2.5)(0,2.25)
\psline{->}(2,2.5)(2,2.25)
\psarc(1,2){1}{180}{360}
\pscircle[linewidth=0.4pt,fillstyle=solid,fillcolor=black](1,1){0.2}
\rput(1,0.25){eval}
\endpspicture
\hskip 2truecm
\pspicture[0.5](0,0)(1,1.5)
\psset{linewidth=\pstlw,xunit=0.5,yunit=0.5,runit=0.5}
\psline(0,3)(0,2)
\psline(2,3)(2,2)
\psline{->}(0,2.5)(0,2.25)
\psline{<-}(2,2.5)(2,2.25)
\psarc(1,2){1}{180}{360}
\pscircle[linewidth=0.4pt,fillstyle=solid,fillcolor=black](1,1){0.2}
\rput(1,0.25){eval}
\endpspicture
\end{align}

\subsection{Product co-product duality (by evaluation)}

The evaluation map provides a natural (functorial) connection 
of products and co-products on $A$ and $A^*$. The action of a linear 
form $\omega$ on a product $m(a \otimes b)$ shall be rewritten as the
sum of scalar products of actions of some tensor 
$\omega_{(1)}\otimes\omega_{(2)}$
on the argument $a\otimes b$ of $m$. In the tangle picture this means that
one pulls the product from the two right down-strands to the single left 
up-strand. During this process the product tangle gets mirrored (rotated 
by $\pi$) and turns into a coproduct tangle acting on the dual space.
The right equation dualizes a product on $A^* \otimes A^*$.
\index{product coproduct duality}
\index{duality!between product and coproduct}
\index{tangle!of product coproduct duality}
\begin{align}
\label{eqn:prcopr}
\pspicture[0.5](0,-0.25)(1.5,2.75)
\psset{linewidth=\pstlw,xunit=0.5,yunit=0.5,runit=0.5}
\psline(0,5)(0,1)
\psline(1,5)(1,2)
\psline(3,5)(3,2)
\psline{<-}(0,4)(0,3.75)
\psline{->}(1,4)(1,3.75)
\psline{->}(3,4)(3,3.75)
\psarc(1,1){1}{180}{360}
\psarc(2,2){1}{180}{360}
\pscircle[linewidth=0.4pt,fillstyle=solid,fillcolor=white](2,1){0.2}
\pscircle[linewidth=0.4pt,fillstyle=solid,fillcolor=black](1,0){0.2}
\endpspicture
\quad=\quad
\pspicture[0.5](0,-0.25)(3,2.75)
\psset{linewidth=\pstlw,xunit=0.5,yunit=0.5,runit=0.5}
\psline(1,5)(1,4)
\psline(4,5)(4,3)
\psline(6,5)(6,2)
\psline(0,3)(0,2)
\psline{<-}(1,4.9)(1,4.75)
\psline{->}(4,4.5)(4,4.25)
\psline{->}(6,4.5)(6,4.25)
\psline(2,0)(4,0)
\psarc(1,3){1}{0}{180}
\psarc(3,3){1}{180}{360}
\psarc(2,2){2}{180}{270}
\psarc(4,2){2}{270}{360}
\pscircle[linewidth=0.4pt,fillstyle=solid,fillcolor=white](1,4){0.2}
\pscircle[linewidth=0.4pt,fillstyle=solid,fillcolor=black](3,0){0.2}
\pscircle[linewidth=0.4pt,fillstyle=solid,fillcolor=black](3,2){0.2}
\endpspicture
\qquad
\pspicture[0.5](0,-0.25)(1.5,2.75)
\psset{linewidth=\pstlw,xunit=0.5,yunit=0.5,runit=0.5}
\psline(3,5)(3,1)
\psline(2,5)(2,2)
\psline(0,5)(0,2)
\psline{->}(3,4)(3,3.75)
\psline{<-}(2,4)(2,3.75)
\psline{<-}(0,4)(0,3.75)
\psarc(2,1){1}{180}{360}
\psarc(1,2){1}{180}{360}
\pscircle[linewidth=0.4pt,fillstyle=solid,fillcolor=white](1,1){0.2}
\pscircle[linewidth=0.4pt,fillstyle=solid,fillcolor=black](2,0){0.2}
\endpspicture
\quad=\quad
\pspicture[0.5](0,-0.25)(3,2.75)
\psset{linewidth=\pstlw,xunit=0.5,yunit=0.5,runit=0.5}
\psline(5,5)(5,4)
\psline(2,5)(2,3)
\psline(0,5)(0,2)
\psline(6,3)(6,2)
\psline(2,0)(4,0)
\psline{<-}(5,4.25)(5,4.4)
\psline{<-}(2,4.5)(2,4.25)
\psline{<-}(0,4.5)(0,4.25)
\psline(4,0)(4,0)
\psarc(5,3){1}{0}{180}
\psarc(3,3){1}{180}{360}
\psarc(2,2){2}{180}{270}
\psarc(4,2){2}{270}{360}
\pscircle[linewidth=0.4pt,fillstyle=solid,fillcolor=white](5,4){0.2}
\pscircle[linewidth=0.4pt,fillstyle=solid,fillcolor=black](3,0){0.2}
\pscircle[linewidth=0.4pt,fillstyle=solid,fillcolor=black](3,2){0.2}
\endpspicture
\end{align}
In terms of algebraic formul\ae\ we can write this as
\begin{align}
&\hskip -0.15truecm
\text{eval}(m(\omega \otimes \omega^\prime) \otimes x) = 
(\text{eval}\otimes \text{eval})(\omega\otimes\omega^\prime\otimes \Delta(x))
= \omega(x_{(2)})\omega^\prime(x_{(1)})\nn
&\hskip -0.15truecm
\text{eval}(\omega \otimes m(x\otimes y)) = 
(\text{eval}\otimes \text{eval})(\Delta(\omega)\otimes x\otimes y)
= \omega_{(1)}(y)\omega_{(2)}(x) 
\end{align}
Using the evaluation map any product induces a coproduct on 
the dual space and vice versa \cite{milnor:moore:1965a}.
\begin{align}
\Delta : A \rightarrow A\otimes A 
\quad \Leftrightarrow \quad
m^* : A^* \otimes A^* \rightarrow A^* \nn
m : A \otimes A \rightarrow A
\quad \Leftrightarrow \quad
\Delta^* : A^* \rightarrow A^* \otimes A^*
\end{align}
Working with a space $A$ and a dual space $A^*$ we are still free to 
choose i) a product and co-product on $A$ {\em or} on $A^*$, ii) 
a product $m$ on $A$ and $m^*$ on $A^*$, i.e. Gra{\ss}mann-Cayley case
or iii) a coproduct $\Delta$  on $A$ and $\Delta^*$ on $A^*$.

\section{Gra{\ss}mann Hopf algebra}

\index{Gra{\ss}mann Hopf algebra}
We will define the Gra{\ss}mann Hopf algebra using the notion
of letters and words, i.e. choosing a basis. Of course one could
reformulate the following results basis free also. However, it will become
important that the structure is unique up to isomorphy only. The notion 
of a Gra{\ss}mann Hopf algebra is standard and may be found 
in \cite{sweedler:1969a}, however we need to introduce some subtleties
which will be used later on and are explained at length in 
\cite{fauser:2002c}. The terms Hopf {\em al}gebra and Hopf gebra denote
in general different structures but coincide in the Gra{\ss}mann Hopf
case, however these terms are distinct e.g. for Clifford Hopf al/gebras.

\index{coproduct!euclidian}
An {\em euclidian coproduct} of a Gra{\ss}mann exterior product $\vee$ 
(meet of hyperplanes) on $A^*$ of an element $x \in A$ is defined as 
the sum over all those tensors $x_{(1)i}\otimes x_{(2)i}$ which multiply 
back to the element $x$. Our terminology reflects the usage of the euclidian 
dual isomorphism $\delta : V \rightarrow V^*$. Hence we consider the 
splits
\begin{align}
I_x \equiv (x) &:= \Big\{\,(a,b) \mid m(a\otimes b)=x\Big\}\nn
& I_{x(1)} \equiv x_{(1)} = a \quad I_{x(2)} \equiv x_{(2)} = b 
\end{align}
and obtain
\begin{align}
\Delta(x) &:= \sum_{I_x}^{\vert I_x\vert} 
I_{x(1)} \otimes I_{x(2)} = \sum_{(x)} x_{(1)} \otimes x_{(2)} 
                        \,=\, x_{(1)} \otimes x_{(2)}\nn
m\circ \Delta(x) &= \sum_{(x)} m(x_{(1)}\otimes x_{(2)}) = 
\vert I_x\vert\ x
\end{align}
The Gra{\ss}mann exterior algebra over a vector space $V^{\wedge}$
having a wedge product $\wedge$ can be obtained by factoring the 
tensor product modulo antisymmetrization. The singe transposition
needed for antisymmetrization is called Gra{\ss}mann crossing or 
{\em graded switch} and is defined as
\index{crossing|see{switch}}
\index{switch!graded}
\begin{align}
\label{eqn:switch}
&\otimes \,\overset{\pi_{\hat{\tau}}}{\longrightarrow}\, \wedge \nn 
&\hat{\tau}(A\otimes B) = (-1)^{\partial A\partial B}B\otimes A
\qquad\text{on homogeneous elements.}
\end{align}
We obtain 
\begin{align}
\Delta_{\wedge}(\Id) &= \Id \otimes \Id \nn
\Delta_{\wedge}(a) &= a \otimes \Id + \Id \otimes a \nn
\Delta_{\wedge}(a \wedge b) &= 
a\wedge b \otimes \Id + a\otimes b - b\otimes a + \Id \otimes a\wedge b\nn
\vdots\nn
\Delta_{\wedge}(x) &= x_{(1)} \otimes x_{(2)}.
\end{align}
The sign stemming from the permutations is included in Sweedler notation.
To establish a basis in a Gra{\ss}mann algebra we need a termordering
\index{termordering!in Gra{\ss}mann basis}
on the elements $a,b,c\ldots \in V$ extended to $V^{\wedge}$ to be able to 
decide if we should solve for $ab$ or $ba=-ab$. The splits of a word 
$A=ab\ldots d$ into two blocks $B=a\ldots c$, $C=b\ldots d$ is such that 
in every block $B,C$ the termordering remains valid. In the Gra{\ss}mann
case we find that a word of length $r$ obeys $2^r$ such splits. The
euclidian dualized wedge coproduct is found to be:
\index{grading!on words and letters}
i) co-unital with counit $\epsilon : V^{\wedge} \rightarrow \openk$,
ii) co-associative, iii) (linear) dual to the exterior product 
(denoted as `vee' $\vee$) on the dual space of linear forms 
$V^{*\,\wedge}$ : $\Delta^* \equiv \vee$ and iv) 
can be {\em obtained in a combinatorial way} by a sum of all `splits`
of the exterior products into 2 blocks.
\index{coproduct!combinatorial properties of}

The pair $(V^{\wedge},\wedge)$ is called {\em Gra{\ss}mann algebra} while
the pair $(V^{\wedge},\Delta_{\wedge})$ is called {\em Gra{\ss}mann cogebra}.
If the coproduct is dualized from teh vee $\vee$ we denote it as 
$\Delta_{\vee}$.
\index{Gra{\ss}mann!algebra}
\index{Gra{\ss}mann!cogebra}

If the following compatibility laws are valid, and if one can
proof that an {\em antipode} exists we can establish a 
{\em Gra{\ss}mann Hopf algebra}.
\index{Gra{\ss}mann Hopf algebra}

In Hopf{\em al}gebras one demands as compatibility laws that product 
and unit are cogebra morphisms and that coproduct and counit are algebra 
morphisms.
\index{Hopf algebra!bihomomorphism axiom} 
\begin{align}
\label{eqn:hom}
\begin{array}{c@{\hskip 0.5truecm}c@{\hskip 0.5truecm}c}
\Rnode{1}{B\otimes B} & \Rnode{2}{B} & \Rnode{3}{B\otimes B} \\[8ex]
\Rnode{4}{B\otimes B\otimes B\otimes B} & & 
\Rnode{5}{B\otimes B\otimes B\otimes B}
\ncline{->}{1}{2}
\Aput{m_B}
\ncline{->}{2}{3}
\Aput{\Delta_B}
\ncline{->}{1}{4}
\Bput{\Delta_B \otimes \Delta_B}
\ncline{->}{4}{5}
\Aput{\Id \otimes \hat{\tau}\otimes \Id}
\ncline{->}{5}{3}
\Aput{m_B \otimes m_B}
\end{array}
\quad
\pspicture[0.5](0,0)(1,2.5)
\psset{linewidth=\pstlw,xunit=0.5,yunit=0.5,runit=0.5}
\psline(0,6)(0,5)
\psline(2,6)(2,5)
\psarc(1,5){1.0}{180}{360}
\psline(1,4)(1,2)
\psarc(1,1){1.0}{0}{180}
\psline(0,1)(0,0)
\psline(2,1)(2,0)
\pscircle[linewidth=0.4pt,fillstyle=solid,fillcolor=white](1,4){0.2}
\pscircle[linewidth=0.4pt,fillstyle=solid,fillcolor=white](1,2){0.2}
\rput(1.0,4.75){$m_B$}
\rput(1.0,1.25){$\Delta_B$}
\endpspicture
\hskip -10pt = \hskip 8pt 
\pspicture[0.5](0,0)(3,2.5)
\psset{linewidth=\pstlw,xunit=0.5,yunit=0.5,runit=0.5}
\psline(1,6)(1,5)
\psline(5,6)(5,5)
\psarc(1,4){1.0}{0}{180}
\psarc(5,4){1.0}{0}{180}
\psline(0,4)(0,2)
\psline(6,4)(6,2)
\psline{c-c}(4,4)(2,2)
\psline[border=4pt,bordercolor=white]{c-c}(2,4)(4,2)
\psarc(1,2){1.0}{180}{360}
\psarc(5,2){1.0}{180}{360}
\psline(1,1)(1,0)
\psline(5,1)(5,0)
\pscircle[linewidth=0.4pt,fillstyle=solid,fillcolor=white](1,1){0.2}
\pscircle[linewidth=0.4pt,fillstyle=solid,fillcolor=white](1,5){0.2}
\pscircle[linewidth=0.4pt,fillstyle=solid,fillcolor=white](5,1){0.2}
\pscircle[linewidth=0.4pt,fillstyle=solid,fillcolor=white](5,5){0.2}
\rput(1.0,4.25){$\Delta_B$}
\rput(1.0,1.75){$m_B$}
\rput(3.0,3.75){$\hat{\tau}$}
\rput(5.0,4.25){$\Delta_B$}
\rput(5.0,1.75){$m_B$}
\endpspicture
\end{align}
Finally we give the axioms for the {\em antipode}, an anti-homomorphism,
and a generalization of the inverse 
\index{antipode!definition}
\begin{align}
S(x_{(1)}) \wedge x_{(2)} = \epsilon(x) \,\Id =
x_{(1)} \wedge S(x_{(2)})  \hskip 2truecm \forall x \in V^{\wedge}
\end{align}
\index{tangle!of antipode}
\begin{align}
\label{eqn:antipode}
\pspicture[0.5](0,0)(1,2.5)
\psset{linewidth=\pstlw,xunit=0.5,yunit=0.5,runit=0.5}
\psline(1,5)(1,4)
\psarc(1,3){1}{0}{180}
\psline(2,3)(2,2)
\psarc(1,2){1}{180}{360}
\psline(1,1)(1,0)
\pscircle[linewidth=0.4pt,fillstyle=solid,fillcolor=white](1,1){0.2}
\pscircle[linewidth=0.4pt,fillstyle=solid,fillcolor=white](1,4){0.2}
\rput(1,1.75){m}
\rput(1,3.25){$\Delta$}
\rput(0,2.5){S}
\endpspicture
\quad&=\quad
\pspicture[0.5](0,0)(1,2.5)
\psset{linewidth=\pstlw,xunit=0.5,yunit=0.5,runit=0.5}
\psline(1,5)(1,3)
\psline(1,0)(1,2)
\pscircle[linewidth=0.4pt,fillstyle=solid,fillcolor=white](1,3){0.2}
\pscircle[linewidth=0.4pt,fillstyle=solid,fillcolor=white](1,2){0.2}
\rput(1.5,3){$\epsilon$}
\rput(1.5,2){$\Id$}
\endpspicture
\quad=\quad
\pspicture[0.5](0,0)(1,2.5)
\psset{linewidth=\pstlw,xunit=0.5,yunit=0.5,runit=0.5}
\psline(1,5)(1,4)
\psarc(1,3){1}{0}{180}
\psline(0,3)(0,2)
\psarc(1,2){1}{180}{360}
\psline(1,1)(1,0)
\pscircle[linewidth=0.4pt,fillstyle=solid,fillcolor=white](1,1){0.2}
\pscircle[linewidth=0.4pt,fillstyle=solid,fillcolor=white](1,4){0.2}
\rput(1,1.75){m}
\rput(1,3.25){$\Delta$}
\rput(2,2.5){S}
\endpspicture
\qquad
\text{U}=\Id\circ\epsilon\,=\,\Id_{\text{conv}} 
\end{align}
A {\em Gra{\ss}mann Hopf algebra} is defined as the following septuple
$H^{\wedge} = (V^{\wedge},\wedge,\Id,$ $\Delta_{\wedge},
\epsilon\,;\hat\tau,S)$ fulfilling the above axioms. A classification 
of convolution algebras obeying a product and a coproduct can be
found in \cite{fauser:2002c}. There it was demonstrated that if
a convolutive unit $\Id_{\text{conv}}$ and an antipode exits then
the product and coproduct induce all other structure tensors in a 
Hopf gebra. This idea goes back to Oziewicz 
\cite{oziewicz:1997a,oziewicz:guzman:2001a}.

The rest of the paper is devoted to the task of showing that almost all
algebraic structures needed in geometry and physics can be {\em derived}
in a plain and natural way from the common generic root of Gra{\ss}mann 
Hopf gebra. In this way we follow Oziewicz \cite{oziewicz:1986a} from 
Gra{\ss}mann to Gra{\ss}mann-Cayley, Clifford, etc. adding in the same 
time the dual structures:\\
{\psset{nodesep=10pt}
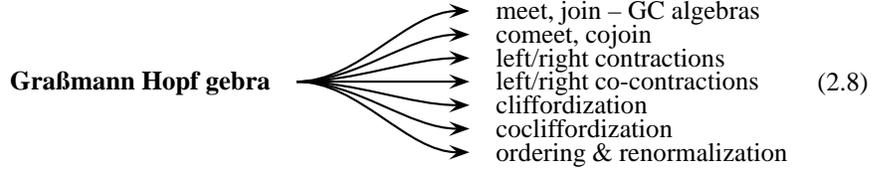
\begin{align}
\hskip -12pt
\begin{tabular}{l@{\hskip 3truecm}l}
 & \Rnode{1}{meet, join -- GC algebras }\\[-3pt]
 & \Rnode{2}{comeet, cojoin}\\[-3pt]
 & \Rnode{3a}{left/right contractions}\\[-3pt]
\Rnode{GHG}{{\bf Gra{\ss}mann Hopf gebra}} & \Rnode{3b}{left/right 
co-contractions}\\[-3pt]
 & \Rnode{4}{cliffordization}\\[-3pt]
 & \Rnode{5}{cocliffordization}\\[-3pt]
 & \Rnode{6}{ordering \& renormalization}
\end{tabular}
\nccurve[angleA=0,angleB=180]{->}{GHG}{1}
\nccurve[angleA=0,angleB=180]{->}{GHG}{2}
\nccurve[angleA=0,angleB=180]{->}{GHG}{3a}
\nccurve[angleA=0,angleB=180]{->}{GHG}{3b}
\nccurve[angleA=0,angleB=180]{->}{GHG}{4}
\nccurve[angleA=0,angleB=180]{->}{GHG}{5}
\nccurve[angleA=0,angleB=180]{->}{GHG}{6}
\end{align}}
We will have no space to discuss the last point here, see 
\cite{brouder:2002a,fauser:2002c,brouder:fauser:frabetti:oeckl:2002a}.

\section{Gra{\ss}mann-Cayley double algebra}
\index{Gra{\ss}mann-Cayley algebra}

\subsection{Integrals and the bracket}

A {\em left (right) integral} is an element $\mu_{L}$ ($\mu_{R}) \in A^*$, 
i.e. a comultivector of the unital cogebra $A^*$ fulfilling:
\index{integral!of a Hopf algebra}
\index{tangle!of integrals}
\begin{align}
\label{eqn:lrintegral}
&\pspicture[0.5](0,0)(1,2.5)
\psset{linewidth=\pstlw,xunit=0.5,yunit=0.5,runit=0.5}
\psline(0,0)(0,2)
\psline(2,1)(2,2)
\psarc(1,2){1}{0}{180}
\psline(1,3)(1,5)
\pscircle[linewidth=0.4pt,fillstyle=solid,fillcolor=white](1,3){0.2}
\pscircle[linewidth=0.4pt,fillstyle=solid,fillcolor=black](2,1){0.2}
\rput(2,0.25){$\mu_R$}
\endpspicture
\quad=\quad
\pspicture[0.5](0,0)(0.5,2.5)
\psset{linewidth=\pstlw,xunit=0.5,yunit=0.5,runit=0.5}
\psline(1,0)(1,2)
\psline(1,3)(1,5)
\pscircle[linewidth=0.4pt,fillstyle=solid,fillcolor=white](1,2){0.2}
\pscircle[linewidth=0.4pt,fillstyle=solid,fillcolor=black](1,3){0.2}
\rput(1.75,2.75){$\mu_R$}
\endpspicture
&&
\pspicture[0.5](0,0)(1,2.5)
\psset{linewidth=\pstlw,xunit=0.5,yunit=0.5,runit=0.5}
\psline(0,1)(0,2)
\psline(2,0)(2,2)
\psarc(1,2){1}{0}{180}
\psline(1,3)(1,5)
\pscircle[linewidth=0.4pt,fillstyle=solid,fillcolor=white](1,3){0.2}
\pscircle[linewidth=0.4pt,fillstyle=solid,fillcolor=black](0,1){0.2}
\rput(0,0.25){$\mu_L$}
\endpspicture
\quad=\quad
\pspicture[0.5](0,0)(0.5,2.5)
\psset{linewidth=\pstlw,xunit=0.5,yunit=0.5,runit=0.5}
\psline(1,0)(1,2)
\psline(1,3)(1,5)
\pscircle[linewidth=0.4pt,fillstyle=solid,fillcolor=white](1,2){0.2}
\pscircle[linewidth=0.4pt,fillstyle=solid,fillcolor=black](1,3){0.2}
\rput(1.75,2.75){$\mu_L$}
\endpspicture
\end{align}
\begin{align}
&(\Id\otimes \mu_R)\Delta(x) = \mu_R(x)\Id
&&(\mu_L\otimes\Id)\Delta(x) = \mu_L(x)\Id \,.
\end{align}
Gra{\ss}mann Hopf gebras are bi-augmented, bi-connected, see
\cite{milnor:moore:1965a,fauser:2002c} and possesses a unique 
left/right integral $\mu$. Integrals in general do {\em not} exist
in Clifford Hopf gebras \cite{fauser:2002c}.

\index{bracket}
The {\em bracket} $[\ldots ]$ of invariant theory is defined to be
a multilinear alternating normalized map of $s$ multivector arguments 
having total degree $n$, i.e. $\partial A_0+\partial A_1 +\ldots 
+\partial A_s = n = \text{dim} V$ and otherwise zero. 
\begin{align}
{}[A_0,...,A_s] &~~:~~ \otimes^s V^{s_i} \longrightarrow \openk \nn
{}[A_0,...,A_s] &~\equiv~ (\mu \circ \wedge^s)(A_0 \otimes \,...\, \otimes A_s)
\end{align}
In fact this is a determinantal map. The unique integral $\mu$ of a 
Gra{\ss}mann Hopf algebra turns out to be the projection onto the 
coefficient of the highest grade element. This allows to define the 
bracket in Gra{\ss}mann Hopf algebraic terms:
\index{tangle!of bracket}
\begin{align}
\label{eqn:bracket}
[A_0,\ldots,A_s]_{\mu} 
\quad&\cong\quad
\pspicture[0.5](0,0)(1.5,2.25)
\psset{linewidth=\pstlw,xunit=0.5,yunit=0.5,runit=0.5}
\psline(1.5,0)(1.5,1)
\psline(0,2.5)(0,3.5)
\psline(3,2.5)(3,3.5)
\psarc(1.5,2.5){1.5}{180}{360}
\psline[linestyle=dotted](1,3)(2,3)
\pscircle[linewidth=0.4pt,fillstyle=solid,fillcolor=white](1.5,1){0.2}
\pscircle[linewidth=0.4pt,fillstyle=solid,fillcolor=black](1.5,0){0.2}
\rput(0,4){$A_0\otimes$}
\rput(1.5,4){$\ldots$}
\rput(3,4){$\otimes A_s$}
\rput(2,0){$\mu$}
\endpspicture
\end{align}
In what follows it is important to realize, that the bracket is 
a sort of {\em cup-tangle} on $n$-strands or equivalently an {\em 
$n \rightarrow 0$ map}. While the evaluation map in (\ref{eqn:eval})
was a pairing of a space and dual space, the bracket, using two arguments,
constitute a self pairing $[.,.] : V^{\wedge}\otimes V^{\wedge} \rightarrow
\openk$. In terms of tangles we can, however, easily transfer 
notions from one to the other case. This will be used in the next 
subsection. 
\index{pairing}
\index{pairing!self}
 
\subsection{Meet and join, linear logic}

\index{meet}
\index{join}
\index{logic!linear}
\index{extensor}
\index{extensor!support of an}
\index{logic!linear XOR}
Let $A$ be an {\em extensor}, i.e. a homogenous decomposable multivector 
which can be written as $A=a_0 \wedge \ldots \wedge a_r$. The linear space 
$\overline{A} = \text{span}\{a_0,\ldots,a_r\}$ is called {\em support of $A$}.
The {\em join ($A \wedge B$)} is defined as the {\em disjoint union} of 
the supports $\overline{A},\,\overline{B}$, i.e.
$\overline{A \wedge B} = \overline{A}\cap\overline{B}$, and zero otherwise
\cite{doubilet:rota:stein:1974a,barnabei:brini:rota:1985a}.
In logical terms this is aa {\em exclusive or (XOR)} on linear spaces.

Geometrically spoken the join connects disjoint geometric elements. Two 
points are joint to span a line, a point and a line may span a plane etc. 
It was already clear to Gra{\ss}mann that one needs a second operation called
{\em meet} (his regressive product, a section) which allows to compute 
common subspaces thereby lowering the degree of the algebraic objects. 
We will show that this notion is natural to a Gra{\ss}mann Hopf algebra.
\index{product!regressive}

{\bf Historical note:}
\index{product!vee -- $\vee$}
\index{Gra{\ss}mann!{\bf A1}}
\index{Gra{\ss}mann!{\bf A2}}
\index{Ergaenzung@Erg\"anzung}
\index{rule of the common factor}
\index{logic!linear NOT}
The {\em meet} or $\vee$-product was introduced by H. Gra{\ss}mann as
'ein\-ge\-wandtes Produkt' in \cite{grassmann:1878a} using what later
was called the {\em rule of the common factor}. He weakened this concept
and renamed the operation to the {\em regressive product} in the 
second Ausdehnungslehre [A2,1862] \cite{grassmann:1862a} using there 
the unary operation of 'Erg\"anzung'. This is the notion of an orthogonal 
complement and was denoted by a vertical line $a \rightarrow \vert a$ such
that $a \wedge \vert a = I$ where $I$ is an element of maximal grade. In 
logical terms this operation is a {\em negation} on a linear space. The 
Erg\"anzung makes explicite use of the total dimension of the underlying 
space $V$ via the element $I$, as it is also well known in logic that negation 
is based on a maximal element in an orthomodular lattice. This Erg\"anzungs
operation of taking the orthogonal complement needs, spoken in geometrical
terms, {\em necessarily} a symmetric polarity which leads necessarily to a 
symmetric polar bilinear form! It is this place where a restriction 
enters. Hence we can address the Erg\"anzung as {\em linear NOT} in
linear logic.

\index{logic!linear de Morgan law}
We call the following rule the {\em de Morgan law for linear spaces}. 
It can be found in Gra{\ss}mann's A2, \cite{grassmann:1862a} and was
reinvented several times, see second line.
\begin{align}
\vert(A\vee B) &= (\vert A) \wedge (\vert B) 
&&\text{[1862, A2]}\nn
A\vee B &= I^{-1}\cdot ((I\cdot A) \wedge (I\cdot B)) &&
\text{\cite{hestenes:sobczyk:1992a} needs a 'dot' product} 
\end{align}
It should be remarked that the usage of a dot- or scalar product is still 
more restrictive than the assumption of orthomodularity which fixes only
a class of polarities having the same determinant.

\index{product!universal formula (Lotze)}
A universal or {\em master formula} for the meet of $r$ factors not using 
any symmetric polarity was given by Alfred Lotze in 1955 \cite{lotze:1955a}%
\footnote{We make use of the definition given by Doubilet, Rota \& Stein 
\cite{doubilet:rota:stein:1974a} which is for two factors, but uses a more
compact notation.}. Lotze showed in a note added in proof of the above 
cited paper, that the meet product turns out to be an exterior product also.
\index{meet!double}
\index{duality!of meet and join}
\index{duality!of projective geometry}
Moreover, Lotze showed that the 'double meet' (meet w.r.t. the meet) is again
probably up to a sign the original wedge product. This is a remarkable
and beautiful duality. Furthermore, it shows that we can safely reject
the idea of Rota to switch the notion of wedge and vee products to come up 
with an direct analogy to set theory since duality spoils a fixed relation. 
Finally this duality shows that it is irrelevant what is a point and what a
hyperplane, but these notions can be interchanged {\em provided} one 
interchanges also the meaning of meet and join. This is the celebrated 
duality of projective geometry.
\myeenv

We are ready to define the meet now entirely in terms of the Gra{\ss}mann 
Hopf algebra as (the signs are due to a reordering of factors):
\index{tangle!of meet}
\begin{align}
A \vee B &:= (a_1\wedge\ldots\wedge a_r) \vee (b_1\wedge\ldots\wedge b_s) \\
&= [B_{(1)},A] \, B_{(2)} =  A_{(1)}\, [B,A_{(2)}] 
= \pm [A,B_{(1)}] \, B_{(2)} =  \pm A_{(1)}\, [A_{(2)},B] \nonumber
\end{align}
The tangle definition of the meet reads:
\begin{align}
\label{eqn:meet}
\pspicture[0.5](0,0)(1,2)
\psset{linewidth=\pstlw,xunit=0.5,yunit=0.5,runit=0.5}
\psline(1,0)(1,1)
\psarc(1,2){1}{180}{360}
\psline(0,2)(0,4)
\psline(2,2)(2,4)
\pscircle[linewidth=0.4pt,fillstyle=solid,fillcolor=white](1,1){0.2}
\rput(1,1.75){$\vee$}
\endpspicture
\quad:=\quad
\pm\,\,\pspicture[0.5](0,0)(2,2)
\psset{linewidth=\pstlw,xunit=0.5,yunit=0.5,runit=0.5}
\psline(0,4)(0,2)
\psline(3,4)(3,3)
\psarc(3,2){1}{0}{180}
\psarc(1,2){1}{180}{360}
\psline(1,1)(1,0.5)
\psline(4,2)(4,0)
\pscircle[linewidth=0.4pt,fillstyle=solid,fillcolor=white](3,3){0.2}
\pscircle[linewidth=0.4pt,fillstyle=solid,fillcolor=white](1,1){0.2}
\pscircle[linewidth=0.4pt,fillstyle=solid,fillcolor=black](1,0.5){0.2}
\rput(1,1.75){$\wedge$}
\rput(3,1.75){$\Delta_\wedge$}
\endpspicture
\quad=\quad
\pm\,\,\pspicture[0.5](0,0)(2,2)
\psset{linewidth=\pstlw,xunit=0.5,yunit=0.5,runit=0.5}
\psline(4,4)(4,2)
\psline(1,4)(1,3)
\psarc(1,2){1}{0}{180}
\psarc(3,2){1}{180}{360}
\psline(3,1)(3,0.5)
\psline(0,2)(0,0)
\pscircle[linewidth=0.4pt,fillstyle=solid,fillcolor=white](1,3){0.2}
\pscircle[linewidth=0.4pt,fillstyle=solid,fillcolor=white](3,1){0.2}
\pscircle[linewidth=0.4pt,fillstyle=solid,fillcolor=black](3,0.5){0.2}
\rput(3,1.75){$\wedge$}
\rput(1,1.75){$\Delta_\wedge$}
\endpspicture
\end{align}
This definition still needs the notion of a maximal grade to exist, 
but works out properly for arbitrary not necessarily symmetric 
non-degenerate bilinear forms too. The meet is a sort of contraction
w.r.t. the self pairing induced by the bracket, see below.

\subsection{Comeet and cojoin}

\index{comeet}
\index{cojoin}
Having the tangle definition it is simply a matter of dualizing to come
up with the notion of a cojoin and comeet. The {\em cojoin} turns out to 
be just the Gra{\ss}mann coproduct $\Delta_{\wedge}$. The {\em co-meet} 
$\Delta_{\vee}$ is given by categorial duality and involves the obvious 
notion of a {\em cointegral}. 
\index{tangle!of comeet}
\begin{align}
\label{eqn:comeet}
\pspicture[0.5](0,0)(1,2)
\psset{linewidth=\pstlw,xunit=0.5,yunit=0.5,runit=0.5}
\psline(1,4)(1,3)
\psarc(1,2){1}{0}{180}
\psline(0,0)(0,2)
\psline(2,0)(2,2)
\pscircle[linewidth=0.4pt,fillstyle=solid,fillcolor=white](1,3){0.2}
\rput(1,2.25){$\Delta_\vee$}
\endpspicture
\quad:=\quad
\pm\,\,\pspicture[0.5](0,0)(2,2)
\psset{linewidth=\pstlw,xunit=0.5,yunit=0.5,runit=0.5}
\psline(0,2)(0,0)
\psline(3,1)(3,0)
\psarc(3,2){1}{180}{360}
\psarc(1,2){1}{0}{180}
\psline(1,3)(1,3.5)
\psline(4,2)(4,4)
\pscircle[linewidth=0.4pt,fillstyle=solid,fillcolor=white](3,1){0.2}
\pscircle[linewidth=0.4pt,fillstyle=solid,fillcolor=white](1,3){0.2}
\pscircle[linewidth=0.4pt,fillstyle=solid,fillcolor=black](1,3.5){0.2}
\rput(3,2.25){$\wedge$}
\rput(1,2.25){$\Delta_\wedge$}
\endpspicture
\quad=\quad
\pm\,\,\pspicture[0.5](0,0)(2,2)
\psset{linewidth=\pstlw,xunit=0.5,yunit=0.5,runit=0.5}
\psline(4,0)(4,2)
\psline(1,0)(1,1)
\psarc(1,2){1}{180}{360}
\psarc(3,2){1}{0}{180}
\psline(3,3)(3,3.5)
\psline(0,2)(0,4)
\pscircle[linewidth=0.4pt,fillstyle=solid,fillcolor=white](1,1){0.2}
\pscircle[linewidth=0.4pt,fillstyle=solid,fillcolor=white](3,3){0.2}
\pscircle[linewidth=0.4pt,fillstyle=solid,fillcolor=black](3,3.5){0.2}
\rput(1,2.25){$\wedge$}
\rput(3,2.25){$\Delta_\wedge$}
\endpspicture
\end{align}
The comeet is a coproduct, i.e. a $1 \rightarrow 2$ map, it may be called
cocontraction w.r.t the cobracket.

\subsection{Gra{\ss}mann-Cayley and fourfold algebra}

\index{Gra{\ss}mann-Cayley algebra}
\index{GC algebra|see{Gra{\ss}mann-Cayley algebra}}
The {\em Gra{\ss}mann-Cayley algebra} is defined to be the
di-algebra $GC(\vee,\wedge)$ having two associative unital binary products.
The various duality relations allow us to identify the Gra{\ss}mann-Cayley
algebra with the Gra{\ss}mann Hopf algebras $H_{\wedge}$ or $H_{\vee}$ over
$V^{\wedge}$ or $V^{*\vee}$ and to introduce a Gra{\ss}mann-Cayley cogebra
$GC(\Delta_{\vee},\Delta_{\wedge})$. In a CD this dualities read as:
\index{tangle!Gra{\ss}mann-Cayley versus Hopf}
\begin{align}
&\begin{array}{ccc}
\Rnode{11}{GC(\vee,\wedge)} &\hskip 2truecm &
\Rnode{12}{GC(\Delta_{\vee},\Delta_{\wedge})} \\[6ex]
\Rnode{21}{H_{\wedge}(\wedge,\Delta_{\vee})} & &
\Rnode{22}{H_{\vee}(\vee,\Delta_{\wedge})} 
\end{array}
\ncline{<->}{11}{12}
\ncline{<->}{11}{21}
\ncline{<->}{12}{22}
\ncline{<->}{21}{22}
\end{align}
Note that in Gra{\ss}mann Hopf algebras the exterior product and the 
exterior coproduct are {\em independent}. This has some subtle
consequences and was the motivation to use wedge and vee for the
exterior products on $V^{\wedge}$ and $V^{*\vee}$, see \cite{fauser:2002c}.
This independence makes it useful to introduce the {\em fourfold algebra}:
\index{algebra!fourfold}
\index{fourfold algebra}
\begin{align}
H_{\wedge}\oplus H_{\vee} &\simeq 
GC(\wedge,\Delta_{\vee},\vee,\Delta_{\wedge}).
\end{align}
It would be interesting to investigate in which way this is a 
Gra{\ss}mann-Cayley Hopf di-algebra. A reasonable assumption is
to relate the wedge $\wedge$ and $\vee$ vee product using an analogy 
of a co-(quasi) triangular structure (which might be trivial),
see \cite{brouder:fauser:frabetti:oeckl:2002a}.

\section{Bilinear forms and contractions}

\subsection{Scalar and coscalar products}

\index{scalar product}
\index{coscalar product}
\index{product!scalar}
\index{product!coscalar}
A {\em scalar product} $B$ on $V\otimes V$ is a map in the set
$\text{lin-hom}(V\otimes V,\openk)$ or similarly on the dual space
$D \in \text{lin-hom}(V^* \otimes V^*,\openk)$ . A {\em coscalar 
product} $C$ is an element of the set $\text{lin-hom}(\openk,V \otimes V)$
or from $\text{lin-hom}(\openk,V^* \otimes V^*)$. 
\vskip 5pt
\begin{align}
\Rnode{1}{V}\hskip 2truecm
\Rnode{2}{V^*}
\ncline[offset=1ex]{->}{1}{2}
\Aput{B}
\ncline[offset=1ex]{->}{2}{1}
\Aput{D}
\qquad
\Rnode{1}{V\otimes V}\hskip 2truecm
\Rnode{2}{\openk}\hskip 2truecm
\Rnode{3}{V^*\otimes V^*}
\ncline[offset=1ex]{->}{1}{2}
\Aput{B}
\ncline[offset=1ex]{->}{2}{1}
\Aput{C}
\ncline[offset=1ex]{<-}{2}{3}
\Aput{D}
\ncline[offset=1ex]{<-}{3}{2}
\Aput{E}
\end{align}
\vskip 5pt
Scalar products are $2\rightarrow 0$ maps, i.e. cup-tangles while coscalar
products are $0 \rightarrow 2$ maps, i.e. cap-tangles. However, on the 
linear spaces $V^{\wedge}$ and $V^{*\vee}$ we have to give a meaning to
a scalar product $B^{\wedge}$, resp. $D^{\vee}$ in a {\em canonical way}. 
Later on we will investigate Clifford algebras where the scalar product is 
the polar bilinear form of a quadratic form on $V$ and the algebra structure 
allows to define a unique generalization.

\index{scalar product!exponentially extended}
If we demand that the scalar product is extended by an exponential map
one can check that this is related to co-(quasi) triangular structures.
Furthermore can show that only exponentially generated scalar products 
$B^{\wedge}$ on $V^{\wedge}\otimes V^{\wedge}$ come up with associative 
algebraic structures during a deformation process
\cite{brouder:2002a,fauser:2002c,brouder:fauser:frabetti:oeckl:2002a}.

The cup-tangles for scalar an coscalar products can be looked at in two 
ways, either as scalar products or as duality in $\text{lin-hom}(V^{\wedge},
V^{*\vee})$ resp. $\text{lin-hom}(V^{*\vee},V^{\wedge})$. This reads:
\index{tangle!of scalar product}
\index{tangle!of coscalar product}
\begin{align}
\label{eqn:cupcap}
\pspicture[0.5](0,0)(1,2)
\psset{linewidth=\pstlw,xunit=0.5,yunit=0.5,runit=0.5}
\psline(0,4)(0,2)
\psline(2,4)(2,2)
\psline{->}(0,2.5)(0,2.25)
\psline{->}(2,2.5)(2,2.25)
\psarc(1,2){1}{180}{360}
\pscircle[linewidth=0.4pt,fillstyle=solid,fillcolor=black](1,1){0.2}
\rput(1,0.25){B}
\endpspicture
\quad\cong\quad
\pspicture[0.5](0,0)(1,2)
\psset{linewidth=\pstlw,xunit=0.5,yunit=0.5,runit=0.5}
\psline(0,4)(0,2)
\psline(2,4)(2,2)
\psline{->}(0,2.5)(0,2.25)
\psline{->}(2,3.7)(2,3.5)
\psline{<-}(2,2.5)(2,2.25)
\psarc(1,2){1}{180}{360}
\pscircle[linewidth=0.4pt,fillstyle=solid,fillcolor=black](1,1){0.2}
\pscircle[linewidth=0.4pt,fillstyle=solid,fillcolor=black](2,3){0.2}
\rput(1.25,3){B}
\rput(1,0.25){eval}
\endpspicture  
\qquad&&~
\pspicture[0.5](0,0)(1,2)
\psset{linewidth=\pstlw,xunit=0.5,yunit=0.5,runit=0.5}
\psline(0,4)(0,2)
\psline(2,4)(2,2)
\psline{<-}(0,2.5)(0,2.25)
\psline{<-}(2,2.5)(2,2.25)
\psarc(1,2){1}{180}{360}
\pscircle[linewidth=0.4pt,fillstyle=solid,fillcolor=black](1,1){0.2}
\rput(1,0.25){D}
\endpspicture
\quad\quad\text{dualized:}\quad
\pspicture[0.5](0,-0.5)(1,2)
\psset{linewidth=\pstlw,xunit=0.5,yunit=0.5,runit=0.5}
\psarc(1,1){1}{0}{180}
\psline(0,1)(0,0)
\psline(2,1)(2,0)
\psline{->}(0,0.5)(0,0.25)
\psline{->}(2,0.5)(2,0.25)
\pscircle[linewidth=0.4pt,fillstyle=solid,fillcolor=black](1,2){0.2}
\rput(1,2.75){C}
\endpspicture
\end{align}
\index{scalar product!canonically induced}
Hence we define the {\em canonically induced scalar product} $B^{\wedge}$,
which fulfils the axioms of a co-(quasi) triangular structure, as:
\begin{align}
\label{eqn:expB}
B^{\wedge} = &\exp_{\wedge}(B) =
\epsilon\otimes\epsilon + B_{ij}\epsilon^i\otimes \epsilon^j
+ B_{[i_1i_2],[j_1j_2]} 
\epsilon^{i_1}\wedge\epsilon^{i_2} \otimes \epsilon^{j_1}\wedge\epsilon^{j_2}
+ \ldots 
\nn
\pspicture[0.5](0,0)(1,2)
\psset{linewidth=\pstlw,xunit=0.5,yunit=0.5,runit=0.5}
\psset{arrowsize=2pt 2,arrowinset=0.2}
\psline(0,4)(0,1)
\psline(2,4)(2,1)
\psline{->}(0,4)(0,3.5)
\psline{->}(2,4)(2,3.5)
\psarc(1,1){1}{180}{360}
\pscircle[linewidth=0.4pt,fillstyle=solid,fillcolor=black](1,0){0.2}
\rput(1,0.75){$B^{\wedge}$}
\endpspicture
\,&=\,
\pspicture[0.5](0,0)(1,2)
\psset{linewidth=\pstlw,xunit=0.5,yunit=0.5,runit=0.5}
\psset{arrowsize=2pt 2,arrowinset=0.2}
\psline(0,4)(0,0)
\psline(2,4)(2,0)
\psline{->}(0,4)(0,3.5)
\psline{->}(2,4)(2,3.5)
\pscircle[linewidth=0.4pt,fillstyle=solid,fillcolor=white](0,0){0.2}
\pscircle[linewidth=0.4pt,fillstyle=solid,fillcolor=white](2,0){0.2}
\endpspicture
\,\oplus\,
\pspicture[0.5](0,0)(1,2)
\psset{linewidth=\pstlw,xunit=0.5,yunit=0.5,runit=0.5}
\psset{arrowsize=2pt 2,arrowinset=0.2}
\psline(0,4)(0,1)
\psline(2,4)(2,1)
\psline{->}(0,4)(0,3.5)
\psline{->}(2,4)(2,3.5)
\psarc(1,1){1}{180}{360}
\pscircle[linewidth=0.4pt,fillstyle=solid,fillcolor=black](1,0){0.2}
\rput(1,0.75){$B$}
\endpspicture
\,\oplus\,
\frac{1}{2!}\,
\pspicture[0.5](0,0)(3,2.5)
\psset{linewidth=\pstlw,xunit=0.5,yunit=0.5,runit=0.5}
\psset{arrowsize=2pt 2,arrowinset=0.2}
\psline(1,5)(1,4)
\psline(5,5)(5,4)
\psline{->}(1,5)(1,4.5)
\psline{->}(5,5)(5,4.5)
\psarc(1,3){1}{0}{180}
\psarc(5,3){1}{0}{180}
\psbezier(2,3)(2,1.5)(2,1.5)(3,1.5)
\psbezier(4,3)(4,1.5)(4,1.5)(3,1.5)
\psbezier(0,3)(0,0)(0,0)(3,0)
\psbezier(6,3)(6,0)(6,0)(3,0)
\pscircle[linewidth=0.4pt,fillstyle=solid,fillcolor=white](1,4){0.2}
\pscircle[linewidth=0.4pt,fillstyle=solid,fillcolor=white](5,4){0.2}
\pscircle[linewidth=0.4pt,fillstyle=solid,fillcolor=black](3,1.5){0.2}
\pscircle[linewidth=0.4pt,fillstyle=solid,fillcolor=black](3,0){0.2}
\rput(3,0.75){$B$}
\rput(3,2.25){$B$}
\endpspicture
\,\oplus\,
\frac{1}{3!}\,
\pspicture[0.5](0,0)(4,3)
\psset{linewidth=\pstlw,xunit=0.5,yunit=0.5,runit=0.5}
\psset{arrowsize=2pt 2,arrowinset=0.2}
\psline(2,6)(2,5)
\psline(6,6)(6,5)
\psline{->}(2,6)(2,5.5)
\psline{->}(6,6)(6,5.5)
\psarc(2,4){1}{0}{180}
\psarc(1,3){1}{0}{180}
\psarc(6,4){1}{0}{180}
\psarc(7,3){1}{0}{180}
\psbezier(3,4)(3,3)(3,3)(4,3)
\psbezier(5,4)(5,3)(5,3)(4,3)
\psbezier(2,3)(2,1.5)(2,1.5)(4,1.5)
\psbezier(6,3)(6,1.5)(6,1.5)(4,1.5)
\psbezier(0,3)(0,0)(0,0)(4,0)
\psbezier(8,3)(8,0)(8,0)(4,0)
\pscircle[linewidth=0.4pt,fillstyle=solid,fillcolor=white](1,4){0.2}
\pscircle[linewidth=0.4pt,fillstyle=solid,fillcolor=white](2,5){0.2}
\pscircle[linewidth=0.4pt,fillstyle=solid,fillcolor=white](6,5){0.2}
\pscircle[linewidth=0.4pt,fillstyle=solid,fillcolor=white](7,4){0.2}
\pscircle[linewidth=0.4pt,fillstyle=solid,fillcolor=black](4,3){0.2}
\pscircle[linewidth=0.4pt,fillstyle=solid,fillcolor=black](4,1.5){0.2}
\pscircle[linewidth=0.4pt,fillstyle=solid,fillcolor=black](4,0){0.2}
\rput(4,0.75){$B$}
\rput(4,2.25){$B$}
\rput(4,3.75){$B$}
\endpspicture
\quad\ldots
\end{align}
The coscalar product $C^{\Delta_{\vee}}$ is obtained in the same way
by categorial duality, i.e. mirroring the tangle horizontally (rotating 
by $\pi$).

\subsection{Contractions}
 
Using the scalar product $B^{\wedge}$ as cup-tangle, we can once
more exploit product coproduct duality. This time all input spaces are
of the same type and we get
\begin{align}
\pspicture[0.5](0,0)(1.5,2.5)
\psset{linewidth=\pstlw,xunit=0.5,yunit=0.5,runit=0.5}
\psset{arrowsize=2pt 2,arrowinset=0.2}
\psline(0,5)(0,1)
\psline(1,5)(1,2)
\psline(3,5)(3,2)
\psline{->}(0,4)(0,3.75)
\psline{->}(1,4)(1,3.75)
\psline{->}(3,4)(3,3.75)
\psarc(1,1){1}{180}{360}
\psarc(2,2){1}{180}{360}
\pscircle[linewidth=0.4pt,fillstyle=solid,fillcolor=white](2,1){0.2}
\pscircle[linewidth=0.4pt,fillstyle=solid,fillcolor=black](1,0){0.2}
\rput(1,0.75){B}
\rput(2,1.75){$\wedge$}
\endpspicture
\quad=\quad
\pspicture[0.5](0,0)(3,2.5)
\psset{linewidth=\pstlw,xunit=0.5,yunit=0.5,runit=0.5}
\psset{arrowsize=2pt 2,arrowinset=0.2}
\psline(1,5)(1,4)
\psline(4,5)(4,3)
\psline(6,5)(6,2)
\psline(0,3)(0,2)
\psline{->}(1,4.5)(1,4.25)
\psline{->}(4,4.5)(4,4.25)
\psline{->}(6,4.5)(6,4.25)
\psline(2,0)(4,0)
\psarc(1,3){1}{0}{180}
\psarc(3,3){1}{180}{360}
\psarc(2,2){2}{180}{270}
\psarc(4,2){2}{270}{360}
\pscircle[linewidth=0.4pt,fillstyle=solid,fillcolor=white](1,4){0.2}
\pscircle[linewidth=0.4pt,fillstyle=solid,fillcolor=black](3,0){0.2}
\pscircle[linewidth=0.4pt,fillstyle=solid,fillcolor=black](3,2){0.2}
\rput(3,0.75){B}
\rput(3,2.75){B}
\endpspicture
\quad=:\quad
\pspicture[0.5](0,0)(1.5,2.5)
\psset{linewidth=\pstlw,xunit=0.5,yunit=0.5,runit=0.5}
\psset{arrowsize=2pt 2,arrowinset=0.2}
\psline(3,5)(3,1)
\psline(2,5)(2,2)
\psline(0,5)(0,2)
\psline{->}(3,4)(3,3.75)
\psline{->}(2,4)(2,3.75)
\psline{->}(0,4)(0,3.75)
\psarc(2,1){1}{180}{360}
\psarc(1,2){1}{180}{360}
\pscircle[linewidth=0.4pt,fillstyle=solid,fillcolor=white](1,1){0.2}
\pscircle[linewidth=0.4pt,fillstyle=solid,fillcolor=black](2,0){0.2}
\rput(2,0.75){B}
\rput(1,1.75){$\LL_B$}
\endpspicture
\end{align}
This motivates the definition of the {\em right contraction} in terms
of a tangle equation as 
\index{contraction!right}
\index{tangle!of right contraction}
\begin{align}
\label{eqn:rightcon}
\pspicture[0.5](0,0)(1,2)
\psset{linewidth=\pstlw,xunit=0.5,yunit=0.5,runit=0.5}
\psset{arrowsize=2pt 2,arrowinset=0.2}
\psline(0,4)(0,2)
\psline(2,4)(2,2)
\psline(1,1)(1,0)
\psarc(1,2){1}{180}{360}
\pscircle[linewidth=0.4pt,fillstyle=solid,fillcolor=white](1,1){0.2}
\rput(1,1.75){$\LL_B$}
\endpspicture
\quad:=\quad
\pspicture[0.5](0,0)(2,2)
\psset{linewidth=\pstlw,xunit=0.5,yunit=0.5,runit=0.5}
\psset{arrowsize=2pt 2,arrowinset=0.2}
\psline(1,4)(1,3)
\psline(0,2)(0,0)
\psline(4,4)(4,2)
\psarc(1,2){1}{0}{180}
\psarc(3,2){1}{180}{360}
\pscircle[linewidth=0.4pt,fillstyle=solid,fillcolor=white](1,3){0.2}
\pscircle[linewidth=0.4pt,fillstyle=solid,fillcolor=black](3,1){0.2}
\rput(3,0.25){$B$}
\rput(1,2){$\Delta^{\wedge}$}
\endpspicture
\end{align}
Moving the product from left to right in the product coproduct
duality (\ref{eqn:prcopr}) gives
\begin{align}
\pspicture[0.5](0,0)(1.5,2.5)
\psset{linewidth=\pstlw,xunit=0.5,yunit=0.5,runit=0.5}
\psset{arrowsize=2pt 2,arrowinset=0.2}
\psline(3,5)(3,1)
\psline(2,5)(2,2)
\psline(0,5)(0,2)
\psline{->}(3,4)(3,3.75)
\psline{->}(2,4)(2,3.75)
\psline{->}(0,4)(0,3.75)
\psarc(2,1){1}{180}{360}
\psarc(1,2){1}{180}{360}
\pscircle[linewidth=0.4pt,fillstyle=solid,fillcolor=white](1,1){0.2}
\pscircle[linewidth=0.4pt,fillstyle=solid,fillcolor=black](2,0){0.2}
\rput(1,2){$\wedge$}
\rput(2,0.75){B}
\endpspicture
\quad=\quad
\pspicture[0.5](0,0)(3,2.5)
\psset{linewidth=\pstlw,xunit=0.5,yunit=0.5,runit=0.5}
\psset{arrowsize=2pt 2,arrowinset=0.2}
\psline(5,5)(5,4)
\psline(2,5)(2,3)
\psline(0,5)(0,2)
\psline(6,3)(6,2)
\psline(2,0)(4,0)
\psline{->}(5,4.5)(5,4.25)
\psline{->}(2,4.5)(2,4.25)
\psline{->}(0,4.5)(0,4.25)
\psline(4,0)(4,0)
\psarc(5,3){1}{0}{180}
\psarc(3,3){1}{180}{360}
\psarc(2,2){2}{180}{270}
\psarc(4,2){2}{270}{360}
\pscircle[linewidth=0.4pt,fillstyle=solid,fillcolor=white](5,4){0.2}
\pscircle[linewidth=0.4pt,fillstyle=solid,fillcolor=black](3,0){0.2}
\pscircle[linewidth=0.4pt,fillstyle=solid,fillcolor=black](3,2){0.2}
\rput(3,0.75){B}
\rput(3,2.75){B}
\endpspicture
\quad=:\quad
\pspicture[0.5](0,0)(1.5,2.5)
\psset{linewidth=\pstlw,xunit=0.5,yunit=0.5,runit=0.5}
\psset{arrowsize=2pt 2,arrowinset=0.2}
\psline(0,5)(0,1)
\psline(1,5)(1,2)
\psline(3,5)(3,2)
\psline{->}(0,4)(0,3.75)
\psline{->}(1,4)(1,3.75)
\psline{->}(3,4)(3,3.75)
\psarc(1,1){1}{180}{360}
\psarc(2,2){1}{180}{360}
\pscircle[linewidth=0.4pt,fillstyle=solid,fillcolor=white](2,1){0.2}
\pscircle[linewidth=0.4pt,fillstyle=solid,fillcolor=black](1,0){0.2}
\rput(2,2){$\JJ_B$}
\rput(1,0.75){B}
\endpspicture
\end{align}
which motivates the definition of the {\em left contraction}
as:
\index{contraction!left}
\index{tangle!of left contraction}
\begin{align}
\label{eqn:leftcon}
\pspicture[0.5](0,0)(1,2)
\psset{linewidth=\pstlw,xunit=0.5,yunit=0.5,runit=0.5}
\psset{arrowsize=2pt 2,arrowinset=0.2}
\psline(0,4)(0,2)
\psline(2,4)(2,2)
\psline(1,1)(1,0)
\psarc(1,2){1}{180}{360}
\pscircle[linewidth=0.4pt,fillstyle=solid,fillcolor=white](1,1){0.2}
\rput(1,1.75){$\JJ_B$}
\endpspicture
\quad:=\quad
\pspicture[0.5](0,0)(2,2)
\psset{linewidth=\pstlw,xunit=0.5,yunit=0.5,runit=0.5}
\psset{arrowsize=2pt 2,arrowinset=0.2}
\psline(3,4)(3,3)
\psline(4,2)(4,0)
\psline(0,4)(0,2)
\psarc(3,2){1}{0}{180}
\psarc(1,2){1}{180}{360}
\pscircle[linewidth=0.4pt,fillstyle=solid,fillcolor=white](3,3){0.2}
\pscircle[linewidth=0.4pt,fillstyle=solid,fillcolor=black](1,1){0.2}
\rput(1,0.25){$B$}
\rput(3,2){$\Delta^{\wedge}$}
\endpspicture
\end{align}
These two definitions are valid for arbitrary inhomogeneous elements 
of any grade. While in textbooks one finds such a definition using the
pairing e.g. \cite{greub:1976a}, there is no direct constructive rule 
for their evaluation. Since we can directly compute coproducts, our
formul\ae\
\begin{align}
\JJ_{B^{\wedge}}( A \otimes B) &= B^{\wedge}(A,B_{(1)})\, B_{(2)} \nn
\LL_{B^{\wedge}}( A \otimes B) &= A_{(1)} \,B^{\wedge}(A_{(2)},B)
\end{align}
are constructive and free of any grade, homogeneity or decomposability
restrictions.

\subsection{Chevalley formul\ae\ for any grade}

In Ref. \cite{chevalley:1997a} Chevalley introduced a recursive method
to compute the contraction. From the properties of the pairing he derived
the following well known rules for the left contraction. Of course
analogous formul\ae\ hold for right contractions. Let $x,y \in V$ 
and $u,v,w \in V^{\wedge}$ the left contraction obeys 
\index{Chevalley formulae@Chevalley formul\ae}
\index{Leibnitz rule!graded}
\begin{align}
\label{eqn:chevalley}
i)  &&  x \JJ_B\, y &= B(x,y)\,\Id = \epsilon( x \circ y)\,\Id \nn
ii) &&  x\JJ_B\, (u \wedge v) &= (x \JJ_B\, u) \wedge v + 
                       \hat{u} \wedge (x \JJ_B\, v) \nn
iii)&&  u \JJ_B\, (v \JJ_B\, w) &= (u \wedge v) \JJ_B\, w,
\end{align}
where $\hat{u}=(-1)^{\partial u}\,u$ is the grade involution which turns
out to be the antipode of the Gra{\ss}mann Hopf algebra \cite{fauser:2002c}.

To show that the above tangle definition of the left contraction is a 
generalization of Chevalley deformation we have to show that the three 
rules (\ref{eqn:chevalley}i--iii) follow from the tangle definition.
But our main aim is to generalize the Chevalley relations to arbitrary
inhomogeneous algebra elements of any grade.

\mybenv{Theorem:}
The left contraction as defined in (\ref{eqn:leftcon}) for 
arbitrary algebra elements generalizes the Chevalley formul\ae\ 
(\ref{eqn:chevalley}i--iii) and reduces to them for the one-vector 
specialization. The graded crossing $\hat{\tau}$ induces the grade 
involution (antipode) in the graded Leibnitz rule (\ref{eqn:chevalley}ii). 
\myeenv 

\noindent{\bf Proof:} of i):
We compute the defining tangle of the left contraction on two grade one 
elements $a,b \in V$:
\begin{align}
\JJ_B(a\otimes b) &= (B^{\wedge}\otimes \Id)\left(
(\Id\otimes \Delta)(a\otimes b)\right) \nn
&=(B^{\wedge}\otimes \Id)(a\otimes b \otimes Id + a\otimes\Id \otimes b) \nn
&=B(a,b)\Id \,. \\
\hskip 2truecm
\text{recall that~~} &~~~~~~ \Delta(b) = b\otimes \Id + \Id \otimes b 
\hskip 3.75truecm\rule{1ex}{1ex}\nonumber
\end{align}
\noindent
But the tangle definition is now valid for arbitrary elements
\begin{align}
\JJ_B(u\otimes v) &= (B^{\wedge}\otimes \Id)\left(
(\Id\otimes \Delta)(u\otimes v)\right) \nn
&=\Big(B^{\wedge}(u,v_{(1)})\Big) v_{(2)}
\end{align}
where only those terms survive having $\partial u = \partial v_{(1)}$.

\noindent{\bf Proof:} of iii):
We compute using tangles the following equation
\begin{align}
\pspicture[0.5](0,0)(1.5,3.5)
\psset{linewidth=\pstlw,xunit=0.5,yunit=0.5,runit=0.5}
\psset{arrowsize=2pt 2,arrowinset=0.2}
\psline(0,7)(0,5)
\psline(2,7)(2,5)
\psline(3,7)(3,3)
\psline(1,4)(1,3)
\psline(2,2)(2,0)
\psarc(2,3){1}{180}{360}
\psarc(1,5){1}{180}{360}
\pscircle[linewidth=0.4pt,fillstyle=solid,fillcolor=white](1,4){0.2}
\pscircle[linewidth=0.4pt,fillstyle=solid,fillcolor=white](2,2){0.2}
\rput(2,2.75){$\JJ_B$}
\endpspicture
\quad&\overset{a)}{=}\quad
\pspicture[0.5](0,0)(2.5,3.5)
\psset{linewidth=\pstlw,xunit=0.5,yunit=0.5,runit=0.5}
\psset{arrowsize=2pt 2,arrowinset=0.2}
\psline(0,7)(0,5)
\psline(2,7)(2,5)
\psline(4,7)(4,5)
\psline(1,4)(1,3)
\psline(5,4)(5,0)
\psline(3,4)(3,3)
\psarc(4,4){1}{0}{180}
\psarc(2,3){1}{180}{360}
\psarc(1,5){1}{180}{360}
\pscircle[linewidth=0.4pt,fillstyle=solid,fillcolor=white](1,4){0.2}
\pscircle[linewidth=0.4pt,fillstyle=solid,fillcolor=white](4,5){0.2}
\pscircle[linewidth=0.4pt,fillstyle=solid,fillcolor=black](2,2){0.2}
\rput(2,2.75){$B$}
\endpspicture
\quad\overset{b)}{=}\quad
\pspicture[0.5](0,0)(3.5,3.5)
\psset{linewidth=\pstlw,xunit=0.5,yunit=0.5,runit=0.5}
\psset{arrowsize=2pt 2,arrowinset=0.2}
\psline(0,7)(0,3)
\psline(2,7)(2,4)
\psline(6,7)(6,6)
\psline(7,5)(7,0)
\psline(6,4)(6,3)
\psarc(6,5){1}{0}{180}
\psarc(5,4){1}{0}{180}
\psarc(3,4){1}{180}{360}
\psarc(2,3){2}{180}{270}
\psarc(4,3){2}{270}{360}
\psline(2,1)(4,1)
\pscircle[linewidth=0.4pt,fillstyle=solid,fillcolor=white](6,6){0.2}
\pscircle[linewidth=0.4pt,fillstyle=solid,fillcolor=white](5,5){0.2}
\pscircle[linewidth=0.4pt,fillstyle=solid,fillcolor=black](3,3){0.2}
\pscircle[linewidth=0.4pt,fillstyle=solid,fillcolor=black](3,1){0.2}
\rput(3,1.75){$B$}
\rput(3,3.75){$B$}
\endpspicture
\nonumber \\[4ex]
&\overset{c)}{=}\quad
\pspicture[0.5](0,0)(3.5,3.5)
\psset{linewidth=\pstlw,xunit=0.5,yunit=0.5,runit=0.5}
\psset{arrowsize=2pt 2,arrowinset=0.2}
\psline(0,7)(0,3)
\psline(2,7)(2,4)
\psline(5,7)(5,6)
\psline(7,4)(7,0)
\psline(4,5)(4,4)
\psline(5,4)(5,3)
\psarc(5,5){1}{0}{180}
\psarc(6,4){1}{0}{180}
\psarc(3,4){1}{180}{360}
\psarc(2,3){2}{180}{270}
\psarc(3,3){2}{270}{360}
\psline(2,1)(3,1)
\pscircle[linewidth=0.4pt,fillstyle=solid,fillcolor=white](5,6){0.2}
\pscircle[linewidth=0.4pt,fillstyle=solid,fillcolor=white](6,5){0.2}
\pscircle[linewidth=0.4pt,fillstyle=solid,fillcolor=black](3,3){0.2}
\pscircle[linewidth=0.4pt,fillstyle=solid,fillcolor=black](2.5,1){0.2}
\rput(2.5,1.75){$B$}
\rput(3,3.75){$B$}
\endpspicture
\quad\overset{a)}{=}\quad
\pspicture[0.5](0,0)(1.5,3.5)
\psset{linewidth=\pstlw,xunit=0.5,yunit=0.5,runit=0.5}
\psset{arrowsize=2pt 2,arrowinset=0.2}
\psline(3,7)(3,5)
\psline(1,7)(1,5)
\psline(0,7)(0,3)
\psline(2,4)(2,3)
\psline(1,2)(1,0)
\psarc(1,3){1}{180}{360}
\psarc(2,5){1}{180}{360}
\pscircle[linewidth=0.4pt,fillstyle=solid,fillcolor=white](2,4){0.2}
\pscircle[linewidth=0.4pt,fillstyle=solid,fillcolor=white](1,2){0.2}
\rput(1,2.75){$\JJ_B$}
\rput(2,4.75){$\JJ_B$}
\endpspicture
\end{align}
We have used the definition of the contraction (\ref{eqn:leftcon}) in 
a), product coproduct duality (\ref{eqn:prcopr}) in b), coassociativity
(\ref{eqn:coass}) in c). This formula was already valid for any grade.
\myeenv

\noindent{\bf Proof:} of ii): The most complicated case is relation 
(\ref{eqn:chevalley}ii). We compute firstly the tangle equation for 
the general case and prove that the restriction to a one-vector argument
yields the well known graded Leibnitz rule.
\begin{align}
\label{eqn:iii)}
\pspicture[0.5](0,0)(1.5,3.5)
\psset{linewidth=\pstlw,xunit=0.5,yunit=0.5,runit=0.5}
\psset{arrowsize=2pt 2,arrowinset=0.2}
\psline(3,7)(3,5)
\psline(1,7)(1,5)
\psline(0,7)(0,3)
\psline(2,4)(2,3)
\psline(1,2)(1,0)
\psarc(1,3){1}{180}{360}
\psarc(2,5){1}{180}{360}
\pscircle[linewidth=0.4pt,fillstyle=solid,fillcolor=white](2,4){0.2}
\pscircle[linewidth=0.4pt,fillstyle=solid,fillcolor=white](1,2){0.2}
\rput(1,2.75){$\JJ_B$}
\endpspicture
\quad&\overset{a)}{=}\quad
\pspicture[0.5](0,0)(2,3.5)
\psset{linewidth=\pstlw,xunit=0.5,yunit=0.5,runit=0.5}
\psset{arrowsize=2pt 2,arrowinset=0.2}
\psline(0,7)(0,3)
\psline(2,7)(2,6)
\psline(4,7)(4,6)
\psline(3,5)(3,4)
\psline(4,3)(4,0)
\psarc(3,3){1}{0}{180}
\psarc(1,3){1}{180}{360}
\psarc(3,6){1}{180}{360}
\pscircle[linewidth=0.4pt,fillstyle=solid,fillcolor=white](3,5){0.2}
\pscircle[linewidth=0.4pt,fillstyle=solid,fillcolor=white](3,4){0.2}
\pscircle[linewidth=0.4pt,fillstyle=solid,fillcolor=black](1,2){0.2}
\rput(1,2.75){$B$}
\endpspicture
\quad\overset{b)}{=}\quad
\pspicture[0.5](0,0)(3.5,3.5)
\psset{linewidth=\pstlw,xunit=0.5,yunit=0.5,runit=0.5}
\psset{arrowsize=2pt 2,arrowinset=0.2}
\psline(0,7)(0,2)
\psline(1,5)(1,3)
\psline(2,7)(2,6)
\psline(6,7)(6,6)
\psline(6,2)(6,0)
\psline(7,5)(7,3)
\psarc(2,5){1}{0}{180}
\psarc(6,5){1}{0}{180}
\psarc(1,2){1}{180}{360}
\psarc(2,3){1}{180}{360}
\psarc(6,3){1}{180}{360}
\psbezier(5,5)(5,4)(3,4)(3,3)
\psbezier[border=4pt,bordercolor=white](3,5)(3,4)(5,4)(5,3)
\pscircle[linewidth=0.4pt,fillstyle=solid,fillcolor=white](2,6){0.2}
\pscircle[linewidth=0.4pt,fillstyle=solid,fillcolor=white](2,2){0.2}
\pscircle[linewidth=0.4pt,fillstyle=solid,fillcolor=white](6,6){0.2}
\pscircle[linewidth=0.4pt,fillstyle=solid,fillcolor=white](6,2){0.2}
\pscircle[linewidth=0.4pt,fillstyle=solid,fillcolor=black](1,1){0.2}
\rput(1,1.75){$B$}
\endpspicture
\nonumber \\[4ex]
&\overset{c)}{=}\quad
\pspicture[0.5](0,0)(5,3)
\psset{linewidth=\pstlw,xunit=0.5,yunit=0.5,runit=0.5}
\psset{arrowsize=2pt 2,arrowinset=0.2}
\psline(0,4)(0,3)
\psline(1,6)(1,5)
\psline(5,6)(5,5)
\psline(9,6)(9,5)
\psline(10,4)(10,2)
\psline(9,1)(9,0)
\psline(2,1)(4,1)
\psarc(1,4){1}{0}{180}
\psarc(5,4){1}{0}{180}
\psarc(9,4){1}{0}{180}
\psarc(9,2){1}{180}{360}
\psarc(2,3){2}{180}{270}
\psarc(3,4){1}{180}{360}
\psbezier(4,1)(5,1)(6,1)(6,2)
\psbezier(6,2)(6,3)(8,3)(8,4)
\psbezier[border=4pt,bordercolor=white](6,4)(6,3)(8,3)(8,2)
\pscircle[linewidth=0.4pt,fillstyle=solid,fillcolor=white](1,5){0.2}
\pscircle[linewidth=0.4pt,fillstyle=solid,fillcolor=white](5,5){0.2}
\pscircle[linewidth=0.4pt,fillstyle=solid,fillcolor=white](9,5){0.2}
\pscircle[linewidth=0.4pt,fillstyle=solid,fillcolor=white](9,1){0.2}
\pscircle[linewidth=0.4pt,fillstyle=solid,fillcolor=black](3,1){0.2}
\pscircle[linewidth=0.4pt,fillstyle=solid,fillcolor=black](3,3){0.2}
\rput(3,1.75){$B$}
\rput(3,3.75){$B$}
\endpspicture
\quad\overset{d)a)}{=}\quad
\pspicture[0.5](0,0)(3,3)
\psset{linewidth=\pstlw,xunit=0.5,yunit=0.5,runit=0.5}
\psset{arrowsize=2pt 2,arrowinset=0.2}
\psline(1,7)(1,6)
\psline(4,7)(4,5)
\psline(6,7)(6,3)
\psline(4,1)(4,0)
\psline(0,5)(0,2)
\psbezier(0,2)(0,1)(3.5,4)(4,3)
\psline[border=4pt,bordercolor=white](3,4)(3,2)
\psarc(1,5){1}{0}{180}
\psarc(3,5){1}{180}{360}
\psarc(5,3){1}{180}{360}
\psarc(4,2){1}{180}{360}
\pscircle[linewidth=0.4pt,fillstyle=solid,fillcolor=white](1,6){0.2}
\pscircle[linewidth=0.4pt,fillstyle=solid,fillcolor=white](3,4){0.2}
\pscircle[linewidth=0.4pt,fillstyle=solid,fillcolor=white](5,2){0.2}
\pscircle[linewidth=0.4pt,fillstyle=solid,fillcolor=white](4,1){0.2}
\rput(3,4.75){$\JJ_B$}
\rput(5,2.75){$\JJ_B$}
\rput(0.5,1.5){eval}
\rput(3.85,3.6){coev}
\endpspicture
\end{align}
We have used the definition of the contraction (\ref{eqn:leftcon}) in 
a), the compatibility of algebra and cogebra structure (\ref{eqn:hom})
in b), product coproduct duality (\ref{eqn:prcopr}) in c), and the following
property of the crossing (\ref{eqn:cross}) in d). 
\begin{align}
\label{eqn:cross}
B^{cs}\tau^{ab}_{sd} &= \tau^{ca}_{ds}B^{sb} .
\hskip 1.5truecm
\pspicture[0.5](0,0)(1,2)
\psset{linewidth=\pstlw,xunit=0.5,yunit=0.5,runit=0.5}
\psset{arrowsize=2pt 2,arrowinset=0.2}
\psbezier(2,4)(2,3)(0,3)(0,2)
\psbezier[border=4pt,bordercolor=white](0,4)(0,3)(2,3)(2,2)
\psline(2,2)(2,0)
\psline(0,2)(0,0)
\psline{->}(2,1.5)(2,1)
\psline{->}(0,2)(0,1.75)
\psline{->}(0,0)(0,0.25)
\pscircle[linewidth=0.4pt,fillstyle=solid,fillcolor=black](0,1){0.2}
\rput(0.75,1){$B$}
\endpspicture
\quad=\quad
\pspicture[0.5](0,0)(1,2)
\psset{linewidth=\pstlw,xunit=0.5,yunit=0.5,runit=0.5}
\psset{arrowsize=2pt 2,arrowinset=0.2}
\psbezier(2,2)(2,1)(0,1)(0,0)
\psbezier[border=4pt,bordercolor=white](0,2)(0,1)(2,1)(2,0)
\psline(2,4)(2,2)
\psline(0,4)(0,2)
\psline{->}(0,3.5)(0,3)
\psline{<-}(2,3.75)(2,4)
\psline{->}(2,2)(2,2.25)
\pscircle[linewidth=0.4pt,fillstyle=solid,fillcolor=black](2,3){0.2}
\rput(1.25,3){$B$}
\endpspicture
\end{align}
In algebraic terms the above given tangle equation (\ref{eqn:iii)}) reads:
\begin{align}
w\JJ_B (u\wedge v) &= (-1)^{(\vert w_{(1)}\vert\vert w_{(2)}\JJ_B\, u\vert)}
\, (w_{(2)}\JJ_B\, u)\wedge (w_{(1)}\JJ_B\, v)
\end{align}
To finish the proof we reduce the general formula to the case of 
a one vector contraction, i.e. we let  $w\rightarrow a \in V$
\begin{align}
a\JJ_B(u\wedge v) &= (a\JJ_B u)\wedge v + \hat{u}\wedge(a \JJ_B v),
\end{align}
remembering the definition of the graded switch (\ref{eqn:switch}) and 
specializing also to a one vector argument in the fist tensor slot
\begin{align}
\hat{\tau}(a \otimes u) &= (-1)^{\partial a\partial u}(u\otimes a) 
= ((-1)^{\partial u}u) \otimes a \nn
&= \hat{u} \otimes a.
\end{align}
we obtain the well known graded Leibnitz rule (\ref{eqn:chevalley}ii). 
\myeenv

Our calculation shows that the grade involution $\hat{u}$ originates
in the graded switch $\hat{\tau}$. The crossing is thus related to the 
derivation property. This observation has tremendous impact on commutation 
relations. Let
\begin{align}
a_i^\dagger \quad &\Leftrightarrow \quad a_i \wedge \nn
a_i         \quad &\Leftrightarrow \quad a_i \JJ_{\delta}  
\end{align}
an note that equation (\ref{eqn:chevalley}ii) defines then the 
commutation relations of such creation and annihilation operations,
i.e. a CAR algebra.
\begin{align}
&a_i a^\dagger_j \mid \phi\rangle = 
  \langle a_i \mid a^\dagger_j \rangle_{\delta} \mid \phi\rangle
- a^\dagger_j a_i \mid \phi\rangle
\end{align}
we have thus shown that all Chevalley deformation formul\ae\ 
follow from the Gra{\ss}mann Hopf gebra generically. Note that 
our formul\ae\ allow to compute expressions having arbitrary grade
or even being inhomogeneous. This will eventually be explored but
see \cite{fauser:1998a,fauser:2001b}.

\subsection{Left/right cocontractions}

\index{cocontraction!left/right}
Recalling that we had defined cap-tangles from co-scalar products
we can employ dualized product coproduct duality to define left
and right cocontractions. We write the tensor of the coscalar product
as $C^{\wedge}_{(1)} \otimes C^{\wedge}_{(2)}$ and define the
left cocontraction via:
\index{tangle!left cocontraction}
\begin{align}
\label{eqn:leftcocon}
\pspicture[0.5](0,0)(1.5,3)
\psset{linewidth=\pstlw,xunit=0.5,yunit=0.5,runit=0.5}
\psset{arrowsize=2pt 2,arrowinset=0.2}
\psline(0,0)(0,4)
\psline(2,0)(2,4)
\psline(3,0)(3,5)
\psarc(1,4){1}{0}{180}
\psarc(2,5){1}{0}{180}
\pscircle[linewidth=0.4pt,fillstyle=solid,fillcolor=white](1,5){0.2}
\pscircle[linewidth=0.4pt,fillstyle=solid,fillcolor=black](2,6){0.2}
\rput(2,5.25){$C$}
\endpspicture
\quad=\quad
\pspicture[0.5](0,0)(3,3)
\psset{linewidth=\pstlw,xunit=0.5,yunit=0.5,runit=0.5}
\psset{arrowsize=2pt 2,arrowinset=0.2}
\psline(0,0)(0,4)
\psline(2,0)(2,3)
\psline(5,0)(5,2)
\psline(2,6)(4,6)
\psline(6,4)(6,3)
\psarc(3,3){1}{0}{180}
\psarc(5,3){1}{180}{360}
\psarc(2,4){2}{90}{180}
\psarc(4,4){2}{0}{90}
\pscircle[linewidth=0.4pt,fillstyle=solid,fillcolor=white](5,2){0.2}
\pscircle[linewidth=0.4pt,fillstyle=solid,fillcolor=black](3,4){0.2}
\pscircle[linewidth=0.4pt,fillstyle=solid,fillcolor=black](3,6){0.2}
\rput(3,5.25){$C$}
\rput(3,3.25){$C$}
\endpspicture
\quad&&\Rightarrow\quad
\pspicture[0.5](0,0)(1,3)
\psset{linewidth=\pstlw,xunit=0.5,yunit=0.5,runit=0.5}
\psset{arrowsize=2pt 2,arrowinset=0.2}
\psline(0,0)(0,3)
\psline(2,0)(2,3)
\psline(1,4)(1,6)
\psarc(1,3){1}{0}{180}
\pscircle[linewidth=0.4pt,fillstyle=solid,fillcolor=white](1,4){0.2}
\rput(1.1,3){$\Delta_{\JJ_C}$}
\endpspicture
\quad:=\quad
\pspicture[0.5](0,0)(2,3)
\psset{linewidth=\pstlw,xunit=0.5,yunit=0.5,runit=0.5}
\psset{arrowsize=2pt 2,arrowinset=0.2}
\psline(0,0)(0,3)
\psline(3,0)(3,2)
\psline(4,3)(4,6)
\psarc(1,3){1}{0}{180}
\psarc(3,3){1}{180}{360}
\pscircle[linewidth=0.4pt,fillstyle=solid,fillcolor=black](1,4){0.2}
\pscircle[linewidth=0.4pt,fillstyle=solid,fillcolor=white](3,2){0.2}
\rput(1,4.75){$C$}
\endpspicture
\end{align}
The right cocontraction follows from
\index{tangle!right cocontraction}
\begin{align}
\label{eqn:rightcocon}
\pspicture[0.5](0,0)(1.5,3)
\psset{linewidth=\pstlw,xunit=0.5,yunit=0.5,runit=0.5}
\psset{arrowsize=2pt 2,arrowinset=0.2}
\psline(3,0)(3,4)
\psline(1,0)(1,4)
\psline(0,0)(0,5)
\psarc(2,4){1}{0}{180}
\psarc(1,5){1}{0}{180}
\pscircle[linewidth=0.4pt,fillstyle=solid,fillcolor=white](2,5){0.2}
\pscircle[linewidth=0.4pt,fillstyle=solid,fillcolor=black](1,6){0.2}
\rput(1,5.25){$C$}
\endpspicture
\quad=\quad
\pspicture[0.5](0,0)(3,3)
\psset{linewidth=\pstlw,xunit=0.5,yunit=0.5,runit=0.5}
\psset{arrowsize=2pt 2,arrowinset=0.2}
\psline(6,0)(6,4)
\psline(4,0)(4,3)
\psline(1,0)(1,2)
\psline(4,6)(2,6)
\psline(0,4)(0,3)
\psarc(3,3){1}{0}{180}
\psarc(1,3){1}{180}{360}
\psarc(4,4){2}{0}{90}
\psarc(2,4){2}{90}{180}
\pscircle[linewidth=0.4pt,fillstyle=solid,fillcolor=white](1,2){0.2}
\pscircle[linewidth=0.4pt,fillstyle=solid,fillcolor=black](3,4){0.2}
\pscircle[linewidth=0.4pt,fillstyle=solid,fillcolor=black](3,6){0.2}
\rput(3,5.25){$C$}
\rput(3,3.25){$C$}
\endpspicture
\quad&&\Rightarrow\quad
\pspicture[0.5](0,0)(1,3)
\psset{linewidth=\pstlw,xunit=0.5,yunit=0.5,runit=0.5}
\psset{arrowsize=2pt 2,arrowinset=0.2}
\psline(2,0)(2,3)
\psline(0,0)(0,3)
\psline(1,4)(1,6)
\psarc(1,3){1}{0}{180}
\pscircle[linewidth=0.4pt,fillstyle=solid,fillcolor=white](1,4){0.2}
\rput(1.1,3){$\Delta_{\LL_C}$}
\endpspicture
\quad:=\quad
\pspicture[0.5](0,0)(2,3)
\psset{linewidth=\pstlw,xunit=0.5,yunit=0.5,runit=0.5}
\psset{arrowsize=2pt 2,arrowinset=0.2}
\psline(4,0)(4,3)
\psline(1,0)(1,2)
\psline(0,3)(0,6)
\psarc(3,3){1}{0}{180}
\psarc(1,3){1}{180}{360}
\pscircle[linewidth=0.4pt,fillstyle=solid,fillcolor=black](3,4){0.2}
\pscircle[linewidth=0.4pt,fillstyle=solid,fillcolor=white](1,2){0.2}
\rput(3,4.75){$C$}
\endpspicture
\end{align}
In terms of algebraic formul\ae\ we find:
\begin{align}
\Delta_{\JJ_C}(x) &= C^{\wedge}_{(1)} \otimes (C^{\wedge}_{(2)} \wedge x) \nn
\Delta_{\LL_C}(x) &= (x\wedge C^{\wedge}_{(1)})\otimes C^{\wedge}_{(2)}
\end{align}

\subsection{Co-Chevalley formul\ae}

Having an exterior coproduct $\Delta$ and a cocontraction we can write 
down immediately the formul\ae\ of co-Chevalley deformation. Let $x \in V$, 
$u \in V^{\wedge}$ we find
\index{co-Chevalley formul\ae}
\index{co-Leibnitz rule}
\begin{align}
\label{eqn:cochevalley}
i)\qquad   & \Delta_{\JJ_C}(\Id) 
\,=\, C^{\wedge}_{(1)} \otimes C^{\wedge}_{(2)}
\nonumber\\[1ex]
i)^\prime\qquad   & \Delta_{\JJ_C}(u) 
\,=\, C^{\wedge}_{(1)} \otimes C^{\wedge}_{(2)} \wedge u
\nonumber\\[1ex]
ii)\qquad & (\Id\otimes \Delta)\Delta_{\JJ_C}(u)
\,=\, C^{\wedge}_{(1)} \otimes \Delta\big(C^{\wedge}_{(2)} \wedge u\big) 
\nonumber\\[1ex]
& 
\phantom{ (\Id\otimes \Delta)\Delta_{\JJ_C}(u)}
\,=\, (-1)^{\partial u_{(1)}\, \partial C^{\wedge}_{(22)}}\,\, 
C^{\wedge}_{(1)} \otimes 
C^{\wedge}_{(21)} \wedge u_{(1)} \otimes 
C^{\wedge}_{(22)} \wedge u_{(2)} 
\nonumber\\[1ex]
ii)^\prime\qquad & (\Id\otimes \Delta)\Delta_{\JJ_C}(x)
\,=\, 
C^{\wedge}_{(1)} \otimes 
C^{\wedge}_{(21)} \otimes 
C^{\wedge}_{(22)} \wedge x 
\nonumber\\[1ex]
& \phantom{\,=\, (\Id\otimes \Delta)\Delta_{\JJ_C}(x)}+
(-1)^{\partial C^{\wedge}_{(22)}}\, 
C^{\wedge}_{(1)} \otimes 
C^{\wedge}_{(21)} \wedge x \otimes 
C^{\wedge}_{(22)}  
\nonumber\\[1ex]
iii)\qquad & (\Delta\otimes \Id)\Delta_{\JJ_C}(u)
\,=\, 
C^{\wedge}_{(11)} \otimes
C^{\wedge}_{(12)} \otimes
C^{\wedge}_{(2)} \wedge u
\end{align}
where $i)$ is equivalent to the coscalar product, $i)^\prime$ is the 
general left cocontraction, $ii)$ is the general cocontraction
on a coproduct and $ii)^\prime$ the corresponding co-Leibnitz rule
for a one vector argument, while $iii)$ dualizes the general formula
(\ref{eqn:chevalley}iii).
These formul\ae\ are new according to our knowledge.

\section{Deformation and cliffordization}

\subsection{Chevalley deformation, cliffordization}

\index{Chevalley deformation}
\index{cliffordization}
Composing the contraction and the exterior multiplication Chevalley
\cite{chevalley:1997a} defined the Clifford product, denoted here as 
$\&c$, as an element $\gamma_x$ of the endomorphism algebra 
$\text{End}(V^{\wedge})$. Let $x \in V$ and $u \in V^{\wedge}$ he 
defined
\begin{align}
\gamma_x &\in \text{End}(V^{\wedge}) \qquad\qquad 
\gamma ~:~ V \otimes V^{\wedge}\rightarrow V^{\wedge}\nn
\gamma_x u &:= x \JJ_B\, u + x \wedge u
\,=\, x \,\&c\, u
\end{align}
Having now general expressions for the contraction and the wedge product 
at hand, its an easy task to write down a grade free Clifford product.
This formula, i.e. the leftmost tangle in (\ref{eqn:cliffordization}), 
was obtained by Rota and Stein \cite{rota:stein:1994a} using Laplace Hopf 
algebras and has been called `Rota-sausage` for obvious reasons by
Oziewicz \cite{oziewicz:2001b,fauser:ablamowicz:2000c}. This process is
a very general deformation of an algebra and not tied to the Clifford case
only. Rota and Stein coined the term {\em cliffordization} but it may
also be addressed as a Drinfeld twist in certain circumstances. The 
cup-tangle in the deformation is a co-(quasi) triangular structure. 
However, {\em only} our approach makes it explicite that cliffordization 
is nothing but the generalized Chevalley deformation and is directly 
composed from left or right contraction and the exterior product, examine 
therefor middle and right tangle.
\index{Chevalley deformation!generalized}
\index{tangle!Rota-sausage}
\index{tangle!of cliffordization}
\index{product!Clifford}
\begin{align}
\label{eqn:cliffordization}
\pspicture[0.5](0,0)(2.5,3.5)
\psset{linewidth=\pstlw,xunit=0.5,yunit=0.5,runit=0.5}
\psline(1,7)(1,6)
\psline(5,7)(5,6)
\psarc(1,5){1}{0}{180}
\psarc(5,5){1}{0}{180}
\psarc(3,5){1}{180}{360}
\psarc(3,5){3}{180}{360}
\psline(3,2)(3,0)
\pscircle[linewidth=0.4pt,fillstyle=solid,fillcolor=black](3,4){0.2}
\pscircle[linewidth=0.4pt,fillstyle=solid,fillcolor=white](3,2){0.2}
\pscircle[linewidth=0.4pt,fillstyle=solid,fillcolor=white](1,6){0.2}
\pscircle[linewidth=0.4pt,fillstyle=solid,fillcolor=white](5,6){0.2}
\endpspicture
\qquad&=\quad
\pspicture[0.5](0,0)(2,3.5)
\psset{linewidth=\pstlw,xunit=0.5,yunit=0.5,runit=0.5}
\psline(1,7)(1,6)
\psline(4,7)(4,5)
\psarc(1,5){1}{0}{180}
\psarc(3,5){1}{180}{360}
\psarc(1.5,2.5){1.5}{180}{360}
\psline(0,5)(0,2.5)
\psline(3,4)(3,2.5)
\psline(1.5,1)(1.5,0)
\pscircle[linewidth=0.4pt,fillstyle=solid,fillcolor=white](1,6){0.2}
\pscircle[linewidth=0.4pt,fillstyle=solid,fillcolor=white](3,4){0.2}
\pscircle[linewidth=0.4pt,fillstyle=solid,fillcolor=white](1.5,1){0.2}
\rput(3,4.75){$\JJ_B$}
\endpspicture
\quad=\quad
\pspicture[0.5](0,0)(2,3.5)
\psset{linewidth=\pstlw,xunit=0.5,yunit=0.5,runit=0.5}
\psline(3,7)(3,6)
\psline(0,7)(0,5)
\psarc(3,5){1}{0}{180}
\psarc(1,5){1}{180}{360}
\psarc(2.5,2.5){1.5}{180}{360}
\psline(4,5)(4,2.5)
\psline(1,4)(1,2.5)
\psline(2.5,1)(2.5,0)
\pscircle[linewidth=0.4pt,fillstyle=solid,fillcolor=white](3,6){0.2}
\pscircle[linewidth=0.4pt,fillstyle=solid,fillcolor=white](1,4){0.2}
\pscircle[linewidth=0.4pt,fillstyle=solid,fillcolor=white](2.5,1){0.2}
\rput(1,4.75){$\LL_B$}
\endpspicture
\end{align}
Note that this product is no longer graded, since we find
\begin{align}
\&c : V^{\wedge^r}\otimes V^{\wedge^s} &\rightarrow
V^{\wedge^{r+s}}\oplus \ldots \oplus V^{\wedge{\vert r-s\vert}}
\end{align}
but obeys only a filtration. This filtration depends on the chosen 
generators i.e. basis. Since the proof that this deformation comes up 
with the Clifford product is given in \cite{rota:stein:1994a} we give only 
a few examples:
\mybenv{Example:} Product of two one-vectors using (\ref{eqn:cliffordization})
middle and right tangle yields
\begin{align}
a \,\&c\, b &= a_{(1)}\wedge (a_{(2)} \JJ_B b) \nn
&= a \wedge (\Id \JJ_B b) + \Id \wedge (a\JJ_B b) \nn
&= a\wedge b + a\JJ_B b = \gamma_a \, b \nn
a \,\&c\, b &= (a \LL_B b_{(1)}) \wedge  b_{(2)} \nn
&= (a \LL_B \Id) \wedge b + (a\LL b)\wedge \Id \nn
&=a\wedge b + b\LL_B a = \gamma_a \, b . 
\end{align}
\myeenv
\mybenv{Example:} Product of two bivectors using (\ref{eqn:cliffordization})
middle tangle 
\begin{align}
(a\wedge b) \,\&c\, (x\wedge y) &= 
(a\wedge b )\wedge (x\wedge y) + a\wedge (b \JJ_B ((x\wedge y))) \nn
&\phantom{=} -b\wedge (a \JJ_B ((x\wedge y))) 
+ \Id\wedge((a\wedge b) \JJ_B (x\wedge y)) \nn
&=\phantom{+}
 a \wedge (b \wedge (x\wedge y) + b \JJ_B (x\wedge y)) \nn
&\phantom{=} +a \JJ_B  (b \wedge (x\wedge y) + b \JJ_B (x\wedge y)) 
-(a \JJ_B b) (x\wedge y) \nn
&=
\gamma_a (\gamma_b (x\wedge y)) -(a\JJ_B b) (x\wedge y) \nn
&=
(\gamma_a \wedge \gamma_b) (x \wedge y) 
\,=\, \gamma_{a\wedge b}(x\wedge y) \, . 
\end{align}
\myeenv

One can prove that \cite{fauser:2002c}:
\begin{itemize}
\item The Gra{\ss}mann Hopf gebra {\em unit $\Id$} remains to be the 
unit, also denoted as $\Id$, of the Clifford product if $B^{\wedge}$ 
is exponentially generated.
\item The Clifford product is {\em associative} if and only if 
$B^{\wedge}$ is exponentially generated.
\item The counit projects products onto the bilinear form 
$\epsilon( u \,\&c\, v) \,=\, B^{\wedge}(u,v)$ 
$(\,=\, \langle 0\mid u \,\&c\, v \mid 0 \rangle\,)$. This can be 
used as {\em vacuum expectation value} in quantum field theory.
\item If $F^{\wedge}$ is exponentially generated from an antisymmetric
bilinear form $F=-F^T$ then is the deformed product $\&c=\dot{\wedge}$ 
again an {\em exterior product}. 
\item The deformation w.r.t such an $F$ encodes the Wick transformation 
of (fer\-mi\-onic) quantum field theory in Hopf algebraic terms 
\cite{fauser:2001b}. 
\item Exponentially generated bilinear forms fulfil the axioms
of a {\em co-(quasi)-triangular structure.}
\item Clifford Hopf gebras are biaugmented but neither connected nor
coconnected, see \cite{milnor:moore:1965a,fauser:2002c} for definitions.
\end{itemize}
These properties result from considering an arbitrary, not exponentially
generated bilinear form ${\cal B\!F}$ on $V^{\wedge}\otimes V^{\wedge}$
in the cliffordization
\begin{align}
a \circ b &:= {\cal B\!F}(a_{(2)},b_{(1)}) a_{(1)} \wedge b_{(2)}
\end{align}
Examining unit, associativity, etc. yields the above claims, see 
\cite{fauser:2002c}.

\subsection{Co-cliffordization, co-Chevalley deformation}

From the Rota sausage tangle (\ref{eqn:cliffordization}), that is from 
the tangle definition of the Clifford product, we derive by categorial 
duality the cocliffordization and the {\em Clifford coproduct} denoted
as $\Delta_{c}$.
\index{cocliffordization}
\index{co-Chevalley deformation}
\index{tangle!co-Clifford product}
\index{coproduct!Clifford}
\begin{align}
\label{eqn:cocliffpr}
\pspicture[0.5](0,0)(1,3.5)
\psset{linewidth=\pstlw,xunit=0.5,yunit=0.5,runit=0.5}
\psset{arrowsize=2pt 2,arrowinset=0.2}
\psline(0,0)(0,3)
\psline(2,0)(2,3)
\psarc(1,3){1}{0}{180}
\psline(1,4)(1,7)
\pscircle[linewidth=0.4pt,fillstyle=solid,fillcolor=white](1,4){0.2}
\rput(1,3.25){$\Delta_c$}
\endpspicture
\quad&:=\quad
\pspicture[0.5](0,0)(2.5,3.5)
\psset{linewidth=\pstlw,xunit=0.5,yunit=0.5,runit=0.5}
\psset{arrowsize=2pt 2,arrowinset=0.2}
\psline(1,0)(1,1)
\psline(5,0)(5,1)
\psarc(1,2){1}{180}{360}
\psarc(5,2){1}{180}{360}
\psarc(3,2){1}{0}{180}
\psarc(3,2){3}{0}{180}
\psline(3,5)(3,7)
\pscircle[linewidth=0.4pt,fillstyle=solid,fillcolor=black](3,3){0.2}
\pscircle[linewidth=0.4pt,fillstyle=solid,fillcolor=white](3,5){0.2}
\pscircle[linewidth=0.4pt,fillstyle=solid,fillcolor=white](1,1){0.2}
\pscircle[linewidth=0.4pt,fillstyle=solid,fillcolor=white](5,1){0.2}
\rput(1,1.75){$m_A$}
\rput(5,1.75){$m_A$}
\rput(3,2.25){$C$}
\rput(3,4.25){$\Delta$}
\endpspicture
\end{align}
Obviously duality tells us that this coproduct is derived from co-Chevalley
deformation also
\begin{align}
\label{eqn:cocliff}
\pspicture[0.5](0,0)(2.5,3.5)
\psset{linewidth=\pstlw,xunit=0.5,yunit=0.5,runit=0.5}
\psset{arrowsize=2pt 2,arrowinset=0.2}
\psline(1,0)(1,1)
\psline(5,0)(5,1)
\psarc(1,2){1}{180}{360}
\psarc(5,2){1}{180}{360}
\psarc(3,2){1}{0}{180}
\psarc(3,2){3}{0}{180}
\psline(3,5)(3,7)
\pscircle[linewidth=0.4pt,fillstyle=solid,fillcolor=black](3,3){0.2}
\pscircle[linewidth=0.4pt,fillstyle=solid,fillcolor=white](3,5){0.2}
\pscircle[linewidth=0.4pt,fillstyle=solid,fillcolor=white](1,1){0.2}
\pscircle[linewidth=0.4pt,fillstyle=solid,fillcolor=white](5,1){0.2}
\endpspicture
\qquad&=\quad
\pspicture[0.5](0,0)(2,3.5)
\psset{linewidth=\pstlw,xunit=0.5,yunit=0.5,runit=0.5}
\psset{arrowsize=2pt 2,arrowinset=0.2}
\psline(1,0)(1,1)
\psline(4,0)(4,2)
\psarc(1,2){1}{180}{360}
\psarc(3,2){1}{0}{180}
\psarc(1.5,4.5){1.5}{0}{180}
\psline(0,2)(0,4.5)
\psline(3,3)(3,4.5)
\psline(1.5,6)(1.5,7)
\pscircle[linewidth=0.4pt,fillstyle=solid,fillcolor=white](1,1){0.2}
\pscircle[linewidth=0.4pt,fillstyle=solid,fillcolor=white](3,3){0.2}
\pscircle[linewidth=0.4pt,fillstyle=solid,fillcolor=white](1.5,6){0.2}
\rput(3,1.25){$\Delta_{\JJ_C}$}
\endpspicture
\quad=\quad
\pspicture[0.5](0,0)(2,3.5)
\psset{linewidth=\pstlw,xunit=0.5,yunit=0.5,runit=0.5}
\psset{arrowsize=2pt 2,arrowinset=0.2}
\psline(3,0)(3,1)
\psline(0,0)(0,2)
\psarc(3,2){1}{180}{360}
\psarc(1,2){1}{0}{180}
\psarc(2.5,4.5){1.5}{0}{180}
\psline(4,2)(4,4.5)
\psline(1,3)(1,4.5)
\psline(2.5,6)(2.5,7)
\pscircle[linewidth=0.4pt,fillstyle=solid,fillcolor=white](3,1){0.2}
\pscircle[linewidth=0.4pt,fillstyle=solid,fillcolor=white](1,3){0.2}
\pscircle[linewidth=0.4pt,fillstyle=solid,fillcolor=white](2.5,6){0.2}
\rput(1.25,1.25){$\Delta_{\LL_C}$}
\endpspicture
\end{align}
and depends thus on the coscalar product in the same manner as the 
Clifford product depends on the scalar product.

\subsection{Deformation from cochains}

\index{Wick normalordering}
It was shown in \cite{fauser:2001b} that the Wick transformation
of normalordered operator products into (non renormalized) time ordered 
operator products can be given by a cliffordization w.r.t. an 
antisymmetric scalar product $F$ exponentially generalized to $F^{\wedge}$.
This is important since renormalization can then be introduced using the
Epstein-Glaser formalism. There is a hope that this can also be achieved
by a product deformation 
\cite{brouder:2002a,brouder:fauser:frabetti:oeckl:2002a}. 
We will not go into this difficult case, but try to show that the
normal ordering transformation is topologically trivial. Therefore we
show that the antisymmetric exponentially generated bilinear form $F^{\wedge}$
can be derived from a cocycle. For precise definitions see \cite{majid:1995a}.

An {\em $r$-cochain} is defined to be a map ${\sf p} : \otimes^r V^{\wedge}
\rightarrow \openk$. A cochain may act in a convolution product, defined
as $f \star g = m (f\otimes g) \Delta$, like an endomorphism ${\cal P} =
{\sf p} \star \Id = \Id \star {\sf p}$ from 
$V^{\wedge} \overset{{\sf p}}{\rightarrow}V^{\wedge}$ where $\Delta$ is
the Gra{\ss}mann coproduct and $m$ the product in $\openk$. Let 
furthermore $\partial$ be a (group like) co-boundary operator, so 
that $\partial {\sf p}$ is a {\em cocycle}. 

Tailoring a 1-cochain to obtain $F^{\wedge}$ as a particular 2-cocycle for
Wick reordering leads to the following requirements for the special 
cochain ${\sf p}$. Let ${\sf p}(\Id)=1$, ${\sf p}(a)=0$ $\forall a\in V$, 
${\sf p}(a\wedge b) = {\sf p}_{ab} \in \openk$ and expand the cochain 
via the Laplace like property ${\sf p}(u \wedge v \wedge w) = 
 \pm\,{\sf p}(u_{(1)},v) {\sf p}(u_{(2)},w)$ to $V^{\wedge}$. Then
define operators ${\cal P}~~{\sf and}~~{\cal P}^{-1}(x)
: V^{\wedge} \rightarrow V^{\wedge}$
\begin{align}
&&{\cal P}(x) = {\rm\sf p}(x_{(1)})x_{(2)} &&
{\cal P}^{-1}(x) = {\rm\sf p}^{-1}(x_{(1)})x_{(2)}&&
\end{align}
which are assumed to be commutative under convolution, i.e.
\begin{align}
{\cal P} &= {\rm\sf p}\star \Id \,=\, \Id \star {\rm\sf p} \nn
{\cal P}^{-1} &= {\rm\sf p}^{-1}\star \Id \,=\, \Id \star {\rm\sf p}^{-1},
\end{align}
The circle product $\circ^{\sf p}$ is defined as an product 
homomorphic to the exterior wedge product under ${\cal P}$.
\begin{align}
&&{\cal P}(x\circ^{\rm\sf p} y) = {\cal P}(x) \wedge {\cal P}(y),&&
{\cal P}^{-1}(x\wedge y) = {\cal P}^{-1}(x) \circ^{\rm\sf p}{\cal P}^{-1}(y),&&
\end{align}
The 2-cocycle derived from the cochain ${\sf p}$ is a bilinear form denoted 
as $\partial {\sf P}$. It is formally invertible and reads explicite
\begin{align}
\partial {\sf P}(u,v) &= 
{\rm\sf p}(u_{(1)}){\rm\sf p}(v_{(2)})
{\rm\sf p}^{-1}(u_{(2)}\wedge v_{(1)}) \nn
\partial {\sf P}^{-1}(u,v) &= 
{\rm\sf p}^{-1}(u_{(1)}){\rm\sf p}^{-1}(v_{(2)})
{\rm\sf p}(u_{(2)}\wedge v_{(1)})
\end{align}
One needs to show that this bilinear form is i) antisymmetric, 
ii) exponentially generated and iii) that the above homomorphism can be 
rewritten as a cliffordization w.r.t. this bilinear form. In terms of 
tangles one has to prove the '=' in the following tangle equation
\index{tangle!owl}
\begin{align}
\pspicture[0.5](0,0)(1,3.5)
\psset{linewidth=\pstlw,xunit=0.5,yunit=0.5,runit=0.5}
\psset{arrowsize=2pt 2,arrowinset=0.2}
\psline(0,7)(0,4)
\psline(2,7)(2,4)
\psarc(1,4){1}{180}{360}
\psline(1,3)(1,0)
\pscircle[linewidth=0.4pt,fillstyle=solid,fillcolor=white](1,3){0.2}
\rput(1,3.75){$\circ^{\rm\sf p}$}
\endpspicture
\quad:=\quad
\pspicture[0.5](0,0)(2.5,3.5)
\psset{linewidth=\pstlw,xunit=0.5,yunit=0.5,runit=0.5}
\psset{arrowsize=2pt 2,arrowinset=0.2}
\psline(1,7)(1,6)
\psline(4,7)(4,6)
\psarc(1,5){1}{0}{180}
\psarc(4,5){1}{0}{180}
\psarc(2,5){2}{180}{270}
\psarc(3,5){2}{270}{360}
\psline(2.5,3)(2.5,2)
\psline(2,3)(3,3)
\psarc(2.5,1){1}{0}{180}
\psline(3.5,1)(3.5,0)
\pscircle[linewidth=0.4pt,fillstyle=solid,fillcolor=white](1,6){0.2}
\pscircle[linewidth=0.4pt,fillstyle=solid,fillcolor=white](4,6){0.2}
\pscircle[linewidth=0.4pt,fillstyle=solid,fillcolor=white](2.5,3){0.2}
\pscircle[linewidth=0.4pt,fillstyle=solid,fillcolor=white](2.5,2){0.2}
\rput(2,4.25){${\rm\sf p}$}
\rput(3,4.25){${\rm\sf p}$}
\rput(1.5,0.25){${\rm\sf p^{-1}}$}
\endpspicture
\quad=\quad
\pspicture[0.5](0,0)(3.5,3.5)
\psset{linewidth=\pstlw,xunit=0.5,yunit=0.5,runit=0.5}
\psset{arrowsize=2pt 2,arrowinset=0.2}
\psline(1,7)(1,6)
\psline(6,7)(6,6)
\psarc(1,5){1}{0}{180}
\psarc(2,4){1}{0}{180}
\psarc(6,5){1}{0}{180}
\psarc(5,4){1}{0}{180}
\psarc(2,4){1}{180}{270}
\psarc(5,4){1}{270}{360}
\psarc(2,3){2}{180}{270}
\psarc(5,3){2}{270}{360}
\psline(0,5)(0,3)
\psline(7,5)(7,3)
\psline(2,3)(5,3)
\psline(2,1)(5,1)
\psline(3.5,1)(3.5,0)
\pscircle[linewidth=0.4pt,fillstyle=solid,fillcolor=white](1,6){0.2}
\pscircle[linewidth=0.4pt,fillstyle=solid,fillcolor=white](2,5){0.2}
\pscircle[linewidth=0.4pt,fillstyle=solid,fillcolor=white](6,6){0.2}
\pscircle[linewidth=0.4pt,fillstyle=solid,fillcolor=white](5,5){0.2}
\pscircle[linewidth=0.4pt,fillstyle=solid,fillcolor=white](3.5,1){0.2}
\rput(3,3.5){${\rm\sf p}$}
\rput(4,3.5){${\rm\sf p}$}
\rput(3.5,2){${\rm\sf p^{-1}}$}
\psline(3.5,3)(3.5,2.5)
\pscircle[linewidth=0.4pt,fillstyle=solid,fillcolor=white](3.5,3){0.2}
\endpspicture
\end{align}
where the rightmost tangle is called {\em owl tangle}.
This was done in \cite{fauser:2002c}. However, it is well known from 
deformation quantization that not every deformation can be written as
a homomorphism of products. Furthermore, since the bilinear form 
$\partial{\sf P}$ is equivalent to a 2-cocycle we see that both 
products, wedge and $\circ^{\sf p} \equiv \dot\wedge$, are topologically
equivalent. However, the related Hopf algebras are quite different
\cite{fauser:2001b}. While the Gra{\ss}mann Hopf algebra w.r.t. the 
wedge is biconnected, that w.r.t. the dotted wedge $\dot\wedge$ is not.

\section{Outlook}

For lack of place we will not give a summary but want to recall shortly the 
main idea and its further eventual development. Indeed the most striking
feature of our approach is its complete duality between algebra and cogebra
structures. Moreover we might have convinced the reader that cogebra
structures are implicitly used e.g. in determinants, combinatorial 
identities, more explicite in the Gra{\ss}mann-Cayley di-algebra having two
associative products one related to a coproduct on the dual space,
and most strikingly in the Clifford product and the very general 
procedure of cliffordization. It has become clear during the course of our
work that the Gra{\ss}mann Hopf algebra is the core and starting point
to develop systematically almost all algebraic structures and less known
costructures. We might remark at this point that it is possible on a formal
level to perform the same reasoning starting with the symmetric Hopf algebra
and deforming it into Weyl or synonymously symplectic Clifford algebras.
We have no time to show that cliffordization is also computationally
very efficient btu see \cite{ablamowicz:fauser:2002a,ablamowicz:fauser:2002b}.

The most intriguing questions for further research are among others the 
following:\\
i) Can a linear cogebra theory be developed including geometrical meaning
without making recourse to the algebra side of the world?\\
ii) Is there a set of axioms which directly characterizes Clifford
Hopf algebras?\\
iii) What is the link of this bigebraic mathematics to geometry and
physics? We know already, that the deformation has to do with quantization
and the propagator of quantum field theory \cite{fauser:1998a,fauser:2001e},
but this relation should be deepened.\\
iv) Since we deal with alternating multivector fields this structure
is very close to string and M-theory, what is the concrete relation?

Lots more questions could be added, but we will insist in a final 
statement, probably of morally nature. Regarding the present development
one {\em cannot go for the algebra only approach} any longer. We hope
that this chapter will push forward this idea.
\vskip 5pt

\noindent{\bf Acknowledgement:}
I would gratefully thank Prof. Heinz Dehnen and the organizers of the
ICCA 6 for financial support.

\small{
\bibliography{products}
\bibliographystyle{plain}
\def\topsep{0pt}
\def\parsep{0pt plus 5pt minus 1pt}
\def\itemsep{-0.5ex}
}
\small
\vskip 1pc
{\obeylines
\noindent Bertfried Fauser
\noindent Universit\"at Konstanz 
\noindent Fachbereich Physik, Fach M678
\noindent D-78457 Konstanz, Germany
\noindent E-mail: Bertfried.Fauser@uni-konstanz.de
\noindent URL: clifford.physik.uni-konstanz.de/\~{}fauser/
}
\vskip 6pt
\noindent Submitted: \date; Revised: TBA.

\normalsize
\noindent
\tableofcontents
\printindex
\end{document}